\documentclass[11pt,a4paper]{article}
\usepackage{amsmath,amssymb,amsfonts,a4wide,graphicx,bm,times}
\usepackage{mcite}
\makeatletter \let\old@startsection=\@startsection
\renewcommand{\@startsection}[6]
{\old@startsection{#1}{#2}{#3}{#4}{#5}{#6\mathversion{bold}}}
\makeatother

\makeatletter

\@addtoreset{equation}{section}
\makeatother
\let\refOld\ref
\renewcommand{\ref}[1]{(\refOld{#1})}

\def\XXint#1#2#3{{\setbox0=\hbox{$#1{#2#3}{\int}$}
\vcenter{\hbox{$#2#3$}}\kern-.5\wd0}}

\newcommand{\tr}{{\rm tr~}}

\newcommand{\superp}[2]{\genfrac{}{}{0pt}{}{#1}{#2}}

\let\paragraphOld\paragraph
\renewcommand{\paragraph}[1]{\paragraphOld{#1}\ \\

\noindent}

 \def\Disc{\text{Disc }}
 \def\d{\delta}
 
 \def\Re{{\rm Re ~}}
 
 \def\p{\partial}
 
 \def\a{\alpha}
 \def\b{\beta}
 \def\g{\gamma}
 \def\d{\delta}
 \def\e{\epsilon}

 \def\l{\lambda}
 \def\m{\mu}

 \def\s{\sigma}
 \def\t{\tau}

 \def\G{\Gamma}

 \def\L{\Lambda}
 \def\O{\Omega}
 \def\o{\omega }
 
\def\la{\left\langle}
\def\ra{\right\rangle}

\def\hf{\dfrac{1}{2}}

\def\implies{\quad\Rightarrow\quad}

\def\mut{\tilde{\mu}}

\def\sign{\text{sign }}

\def\CF{\mathcal{F}}

\def\CA{{\mathcal{A}}}
\def\CZ{{\mathcal{Z}}}
\def\CB{{\mathcal{B}}}

\def\tmu{\tilde{\mu}}

\def\CZb{\mathcal{Z}_{\b}}
\def\bens{$\b$-ensemble}

\def\hpsi{\hat\psi}

\def\ol{_\text{pert.}}
\def\inst{_\text{inst.}}
\def\full{_{\mathcal{N}=2}}
\def\brho{\bar\rho}

\def\CS{\mathcal{S}}
\def\hphi{\hat\phi}
\def\ct{_{n_l}}
\def\mass{_\text{mass}}
\def\tg{\tilde{\g}}
\def\tC{\tilde{C}}
\def\SYM{_\text{SYM}}

\begin{document}
\begin{titlepage}
\renewcommand{\thefootnote}{\fnsymbol{footnote}}

\vspace*{1cm}
    \begin{Large}
       \begin{center}
         {Large $N$ techniques for Nekrasov partition functions and AGT conjecture}
       \end{center}
    \end{Large}
\vspace{0.7cm}

\begin{center}
Jean-Emile B{\sc ourgine}\footnote
            {
e-mail address : 
jebourgine@sogang.ac.kr}\\
      
\vspace{0.7cm}                    
{\it Department of Physics,
}\\
{\it Sogang University, Seoul 121-742, Korea}
\end{center}

\vspace{0.7cm}

\begin{abstract}
\noindent
The AGT conjecture relates $\mathcal{N}=2$ 4d SUSY gauge theories to 2d CFTs. Matrix model techniques can be used to investigate both sides of this relation. The large $N$ limit refers here to the size of Young tableaux in the expression of the gauge theory partition function. It corresponds to the vanishing of $\O$-background equivariant deformation parameters, and should not be confused with the t'Hooft expansion at large number of colors. In this paper, a saddle point approach is employed to study the Nekrasov-Shatashvili limit of the gauge theory, leading to define $\b$-deformed, or quantized, Seiberg-Witten curve and differential form. Then this formalism is compared to the large $N$ limit of the Dijkgraaf-Vafa \bens. A transformation law relating the wave functions appearing at both sides of the conjecture is proposed. It implies a transformation of the Seiberg-Witten $1$-form in agreement with the definition specified earlier. As a side result, a remarkable property of $\mathcal{N}=2$ theories emerged: the instanton contribution to the partition function can be determined from the perturbative term analysis.
\end{abstract}
\vfill

\end{titlepage}
\vfil\eject

\setcounter{footnote}{0}

\section{Introduction and summary}
A very powerful technique to evaluate matrix model integrals consists in expending correlators and partition function in the matrix size $N$. Solving recursively Schwinger-Dyson, or loop, equations, the Eynard-Orantin topological recursion provides the expression of these quantities at all orders in the $1/N^2$ expansion \cite{Eynard2007a}. The formalism survives the $\b$-deformation \cite{Chekhov2006,*Chekhov2010a,Eynard2008a,*Chekhov2009,*Chekhov2010} and can be applied to the Dijkgraaf-Vafa (DV) \bens\ representing Liouville correlators \cite{Dijkgraaf2009}. It was recently realized that this large $N$ technique can be employed directly within Liouville theory to solve perturbatively a set of Ward identities, leading to the semi-classical expansion of correlation functions \cite{Chekhov2012a}. In the spirit of AGT conjecture \cite{Alday2009}, a similar procedure should be available to treat directly the instanton partition function of four dimensional gauge theories with $\mathcal{N}=2$ supersymmetry.\footnote{The original conjecture deals with gauge groups being a product of $SU(2)$ groups and correlators of Liouville theory. Various verifications of the conjecture can be found in \cite{Mironov2009a,*Giribet2009,*Marshakov2009a,*Nanopoulos2009,*Alba2009,*Marshakov2009b,*Belavin2011,*Kanno2011,Mironov2009}, and a proof was given in the series of papers \cite{Fateev2009,*Hadasz2010,*Alba2010}. The extension to $SU(N_c)$ gauge groups and Toda field theory was proposed and investigated in \cite{Wyllard2009,*Bonelli2009,*Fateev2011,Mironov2010b}. According to Dijkgraaf and Vafa \cite{Dijkgraaf2009}, the conjecture can be reformulated using a $\b$-deformed matrix model. This proposal is discussed in \cite{Mironov2009b,Mironov2010c,Mironov2010e,*Itoyama2010a,*Itoyama2011,*Fujita2009}, and also extends to higher genus and various gauge groups \cite{Itoyama2009,Bonelli2011,*Maruyoshi2011,*Schiappa2009}. The DV formulation of the conjecture has been demonstrated only for the case $\b=1$ which corresponds to the Hermitian matrix model \cite{Mironov2010d,Zhang2011}.}$ ^,$\footnote{The link between the moduli space of $\mathcal{N}=2$ SYM instantons on $\mathbb{R}^4$ and Liouville theory was previously mentioned in \cite{Matone2003,*Bertoldi2004}, hinted by an earlier matrix model proposal \cite{Dijkgraaf2002a}.} Such a technique would find applications in a large set of problems where localization is employed and results in a set of coupled integrals.

The first step in this direction was made ten years prior to AGT by Nekrasov and Okounkov \cite{Nekrasov2003a}. In this work, they studied the partition function of $\mathcal{N}=2$ (matter) $SU(N_c)$ super Yang-Mills theories defined on the $\O$-background. They argued that when both equivariant deformation parameters $\e_1,\e_2$ tend to zero, the sum over Young tableaux involved in the expression of the partition function \cite{Nekrasov2003} is dominated by a partition of diverging size $k\sim 1/(\e_1\e_2)$. This partition is characterized by a continuous density of boxes, the limit shape. Introducing a collective field action for the limit shape, they obtained its expression as a solution of a saddle point equation. They reproduced the Seiberg-Witten (SW) curve \cite{Seiberg1994,*Seiberg1994a}, and showed that the instanton free energy reduces to the SW prepotential. The strength of this technique lies in the possibilities of generalization. It has been used in \cite{Nekrasov2012} to solve the full set of quiver $\mathcal{N}=2$ theories defined over a general (affine) Lie algebra \cite{Gaiotto2009a}.

Since the theory on $\mathbb{R}^4$ is recovered as $\e_1,\e_2\to0$, this limit will be referred to as the SW limit. Following the AGT-DV correspondence, it is equivalent to the large $N$ (or planar) limit of the \bens\ at $\b=1$. Tuning $\b$ with $N$, it is possible to define a double-scaled planar limit for the $\b$-ensemble which corresponds to the Nekrasov-Shatashvili (NS) limit of the $\O$-background, namely $\e_2$ tends to zero while $\e_1$ is fixed \cite{Nekrasov2009}. As in the SW limit, large $N$ techniques can be developed to handle the partition function of $\mathcal{N}=2$ SUSY gauge theories \cite{Poghossian2010,*Fucito2011}. Analysis of quiver theories was recently performed along these lines in \cite{Fucito2012}, extending the work \cite{Nekrasov2012} to the NS background. One of the motivations for the present paper is to provide more details on this construction. To do so, we consider a slightly different point of view from \cite{Poghossian2010}, and take $\e_1$ to be infinitesimal instead of finite. The main reason for this choice is to establish $\b$-deformed SW relations order by order.

The two celebrated formulas derived by Seiberg and Witten in \cite{Seiberg1994} relate Coulomb branch vevs and prepotential to the cycle integrals of a differential form over an algebraic curve. Similar relations hold for the matrix model filling fractions and planar free energy, obtained as cycle integrals of the resolvent over the spectral curve \cite{Chekhov2002}. The $\b$-deformation of these relations is very natural given the correspondence between SW theory and integrable systems \cite{Gorsky1995}. This correspondence was originally observed for gauge theories defined on $\mathbb{R}^4$, and classical integrable models. It was later extended by Nekrasov and Shatashvili to quantum integrable systems and gauge theories defined on the $\Omega$-background with a vanishing parameter $\e_2$ \cite{Nekrasov2009}. The non-zero parameter $\e_1$ plays the role of a Planck constant in the quantized model. We focus here on gauge theories with $SU(N_c)$ gauge group and $N_f$ fundamental hypermultiplets. The associated integrable model is an Heisenberg XXX spin chain with $N_c$ sites \cite{Gorsky1996}. Remarkably, another integrable system shows up in the NS limit on the CFT side of the AGT correspondence. In the case of Liouville theory ($N_c=2$), it is a reduced Gaudin model defined on a punctured sphere and characterized by a Schr\"odinger equation. The induced correspondence between quantum integrable models is called the bispectral duality \cite{Zenkevich2011,*Mironov2012a,Mironov2012b}. The relation between Liouville theory and Gaudin model motivated the introduction in \cite{Mironov2010,*Mironov2010a,*Popolitov2010} of $\b$-deformed (or quantized) SW relations for the DV \bens. It was indeed observed that $\e_1$-corrections to the prepotential can be computed as Bohr-Sommerfeld integrals coming from the semi-classical treatment of the Schr\"odinger problem. These integrals extend the standard SW theory to the NS background. The fact that the \bens\ free energy obeys such deformed SW relations was later established in \cite{Bourgine2012} using the Dyson collective field representation. The use of a semi-classical approach explains the choice of keeping $\e_1$ infinitesimal. At $\e_1$ finite, the algebraic SW curve is no longer well-defined, and the notion of ``quantum curve'' must be used \cite{Eynard2008a}.

Here, our starting point is the difference equation \ref{BaxterTQ}, an analogue of Baxter TQ relation derived in \cite{Poghossian2010} using a large $N$ technique to handle the gauge theory in the NS limit. This equation can also be obtained from the Liouville side's Schr\"odinger equation under a quantum change of variable. It is used to write an effective action for the gauge theory as $\e_2\to0$ which is similar to the \bens 's Dyson action. Reproducing the method employed in \cite{Bourgine2012} to treat the \bens, we derive the $\b$-deformed SW relations. Those are the main result of this paper.

\paragraph{Outline and key results}
This paper is divided into three parts. The purpose of the first section is to develop an effective field theory describing the NS limit. We follow both approaches employed in the papers \cite{Nekrasov2003a} and \cite{Poghossian2010}, leading us to construct two equivalent field theories upon different collective fields. In the work of Poghossian et. al. \cite{Poghossian2010}, the collective field is the instanton density defined in \ref{def_rinst}. From this object, we find more convenient to introduce another density, denoted $\rho(t)$ and defined in the section \refOld{Rev_pert}. This definition is rather subtle as it involves the splitting of $\rho$ into instanton $\rho\inst$ and perturbative $\rho\ol$ densities. In the same way, the effective action constructed from the Nekrasov instanton partition function decomposes into a difference of two terms. The first term describes the full partition function, whereas the second one reproduces the perturbative contribution to the free energy. It is further observed that both terms consist of the same action, evaluated for the densities $\rho$ and $\rho\ol$ respectively. Given $\rho\ol$, it is possible to determine this action knowing only the perturbative part of the partition function. Then, the full partition function can be recovered from a saddle point method. It seems reasonable to assume some universality in the definition of $\rho\ol$ which is remarkably simple. Thus, we conclude that the perturbative contribution to the free energy also determines the instanton corrections. It is an important side result of our paper.

On the other hand, in \cite{Nekrasov2003a} Nekrasov and Okounkov used an average profile of Young tableaux as collective field. This shape function $f(t)$ is defined in \ref{shape}. It naturally appears when the integrals \ref{Zinst} present in Nekrasov's original formula \cite{Nekrasov2003} are evaluated as sum over their residues. Indeed, it is a well-known fact that these residues are in one to one correspondence with Young tableaux boxes. In the NS limit, the two collective fields $\rho$ and $f$ are related through the difference equation,
\begin{equation}\label{fri}
f''(t)=(e^{-\e_1\p_t}-1)\rho(t).
\end{equation}
Using this relation, we derive an effective action for $f$ from the action \ref{SSYM} over $\rho$ obtained previously. In the case of pure SYM, it reads
\begin{equation}\label{CSSYM_f}
\CS\SYM[f'']=-\dfrac1{\e_1}\int_{\mathbb{R}^2}{dtdsf''(t)f''(s)\tg_{\e_1}(t-s)}+\dfrac1{2\e_1}\log q\int_\mathbb{R}{dt\ t^2f''(t)},
\end{equation}
the function $\tg_\e$ being defined in appendix \refOld{AppA}.

Once collective field actions have been obtained, we proceed to study the equation of motion in section three. In order to find agreement with the Baxter equation \ref{BaxterTQ} proposed in \cite{Poghossian2010}, the effective action has to be amended by a cut-off term,
\begin{equation}
\CS\SYM[f'']=-\dfrac1{\e_1}\int_{\mathbb{R}^2}{dtdsf''(t)f''(s)\tg_{\e_1}(t-s)}+\dfrac1{2\e_1}\log q\int_\mathbb{R}{dt\ t^2f''(t)}+\int_\mathbb{R}{l^I(f'(t))dt},
\end{equation}
with the function $l^I(x)$ such that $l^I(x)'=\log|\sin(\pi x)|$. Such a cut-off term was first observed by Dyson in the context of matrix models \cite{Dyson1962}. It originates from the regularization of divergences at coincident eigenvalues in the action, and from the replacement of discrete integrations by a functional integral. This term is responsible for an additional contribution to the SW differential that exhibits poles at the branch points. This contribution can be eliminated at the cost of introducing $\e_1$-corrections to the SW relations, namely
\begin{equation}\label{SW_B}
\dfrac{\p\CF}{\p a_l}=\oint_{B_l}{dS}-i\pi\e_1(l-1),\quad a_l=\dfrac1{2i\pi}\oint_{A_l}{dS}+\dfrac{\e_1}{2},
\end{equation}
with $\CF$ denoting the free energy, $a_l$ the Coulomb branch vevs, and $dS$ the deformed SW differential \ref{def_SW}. The expression of the action cut-off term, and the induced SW relations are our main results.

The similarities between the treatment of the gauge theory partition function presented here and the standard large $N$ approach to matrix models and \bens s makes the comparison between gauge theory and DV model much easier, leading to a very explicit expression of the AGT correspondence. This correspondence is worked out in the last part of the paper for the case of an $SU(2)$ gauge group and four fundamental flavors (SCQCD). It takes the form of a quantum change of variable analogue to the one used by Gaiotto to rewrite the SW curve \cite{Gaiotto2009a}. This change of variable transforms the Schr\"odinger equation of the Gaudin model into the  Baxter relation \ref{BaxterTQ} studied in the preceding section. The AGT derivation of the equation \ref{BaxterTQ} validates the results derived in \cite{Poghossian2010}, providing further motivation to introduce the cut-off term in the effective action. It also implies that wave functions solving Schr\"odinger and Baxter equations are related through a Mellin transform \ref{Mellin}. This proposal is analyzed at first orders in $\e_1$, and consistency with the definition of SW differentials is observed.

\section{Gauge theory partition function in the NS limit}
We consider an $\mathcal{N}=2$ gauge theory with $SU(N_c)$ gauge group, and $N_f$ massive hypermultiplets in the fundamental representation. When the number of flavors happens to be twice the number of colors, the $\b$-function of the gauge coupling $\t$ is vanishing, and the theory is superconformal. The dependence in the coupling constant will be encoded in the parameter $q=e^{2\pi i\t}$. For $N_f\neq 2N_c$, the gauge coupling is renormalized by a UV cutoff which generates an effective scale that we also denote $q$.\footnote{It is related to the parameter $\L_\text{[NO]}$ used in \cite{Nekrasov2003a} by $q=\L_\text{[NO]}^{2N_c-N_f}$.} The theory is considered on the Coulomb branch, and the vev of the adjoint scalar fields are denoted $a_l$ where $l$ runs over the color index. Using a localization technique, the instanton contribution to the partition function is reduced to a sum over coupled contour integrals \cite{Nekrasov2003,Losev1997},
\begin{equation}\label{Zinst}
\CZ\inst(q,\e_1,\e_2,a_l,\mu_r)=1+\sum_{k=1}^\infty{\dfrac{q^k}{k!}\left(\dfrac{\e_1+\e_2}{\e_1\e_2}\right)^k\int_\mathbb{R}{\prod_{I=1}^k{\dfrac{Q(\phi_I)d\phi_I}{2i\pi}}\prod_{\superp{I,J=1}{I\neq J}}^k{D(\phi_I-\phi_J)}}}.
\end{equation}
The function $D$ couples the integrals, it is a simple rational fraction involving only $\e$-parameters,
\begin{equation}
D(t)=\dfrac{t(t+\e_1+\e_2)}{(t+\e_1)(t+\e_2)}.
\end{equation}
The potential $Q$ is a ratio of matter and gauge polynomials,
\begin{equation}
Q(t)=\dfrac{M(t)}{A(t+\e_1+\e_2)A(t)},\quad M(t)=\prod_{r=1}^{N_f}{(t-\m_r)},\quad A(t)=\prod_{l=1}^{N_c}{(t-a_l)},
\end{equation}
where $\m_r$ denotes the mass of the hypermultiplets. The expression \ref{Zinst} is manifestly invariant under the sign flip of the parameters $a_l$, $\mu_r$, $\e_1$ and $\e_2$, together with $q\to(-1)^{N_f}q$.\footnote{A different convention on the sign of the masses $\mu_r$ is used in \cite{Alday2009} and \cite{Nekrasov2003,Nekrasov2003a}. For consistency with the last section, we employ here the convention of AGT \cite{Alday2009}.}

Contour integrals in the expression $\ref{Zinst}$ of the partition function can be evaluated as sums over residues \cite{Nekrasov2003}. Poles on the real line, namely for $\phi_I-\phi_J$ equals to $\e_1$ or $\e_2$, are avoided by a shift of the integration contours equivalent to the replacement $\e_{1,2}\to\e_{1,2}+i0$. A similar treatment is applied to the poles at $\phi_I=a_l$, with the shift $a_l\to a_l+i0$. Contributing poles are interpreted as the instanton positions, they are indexed by the boxes of $N_c$ Young tableaux $Y_l$, and will be denoted $\phi_{l,i,j}$, where $l$ is the color index, and $(i,j)\in Y_l$ parameterizes the position of the box in the $l$th tableau. We further denote $\l_i^{(l)}$ the height of the $i$th column of the Young tableau $Y_l$, $n_l$ its number of columns and $|Y_l|$ the number of boxes. Explicitly, poles are given by
\begin{equation}
\phi_{l,i,j}=a_l+(i-1)\e_1+(j-1)\e_2,
\end{equation}
with the indices $(i,j)$ ranging from $i=1\cdots n_l$ and $j=1\cdots \l_i^{(l)}$. The instanton partition function \ref{Zinst} rewrites 
\begin{equation}\label{sum_poles}
\CZ\inst(q,\e_1,\e_2,a_l,\mu_r)=\sum_{\{Y_1,\cdots,Y_{N_c}\}}{q^k{\prod_{I,J=1}^{k}}{D(\phi_I-\phi_J)}\prod_{I=1}^k{Q(\phi_I)}}.
\end{equation}
Indices in the products decompose into color and box position as $I=(l,i,j)$. They run over the total number $k=\sum_l{|Y_l|}$ of boxes in the partition, and the $q$-expansion is an expansion over the size of partitions. At equal color indices, spurious zeros and divergences may appear within the products of $D(\phi_I-\phi_J)$ and $Q(\phi_I)$. Such factors cancel with each other and the whole product is well defined for any Young tableau. 

\subsection{Effective action for large Young tableaux}
In order to define an effective field theory, a collective field have to be introduced. Following Poghossian et. al. \cite{Poghossian2010,*Fucito2011}, we first consider the density of instantons
\begin{equation}\label{def_rinst}
\brho\inst(\phi)=\dfrac{\e_1\e_2}{\e_1+\e_2}\sum_{I=1}^{k}{\d(\phi-\phi_I)}.
\end{equation}
Let us examine how this object behaves in the NS limit. As $\e_2\to0$, the size of Young tableaux diverges, and the height of each column $\l_i^{(l)}$ is sent to infinity such that $\e_2\l_i^{(l)}$ remains finite. The row index $\e_2(j-1)$ becomes a continuous variable in the interval $[0,\e_2\l_i^{(l)}]$, whereas the column index $i$ remains discrete. Thus, the instanton positions $\phi_{l,i,j}$ are now continuous variables $\phi_{l,i}$ that belong to the intervals $[t_{l,i}^0,t_{l,i}]$ where $t_{l,i}^0=a_l+(i-1)\e_1$ and $t_{l,i}=t_{l,i}^0+\l_i^{(l)}\e_2$. In this process, an integral substitutes the sum over the index $j$, and for any function $F(\phi)$,
\begin{equation}\label{limit_NS}
\e_2\sum_{(i,j)\in Y_l}{F(\phi_{l,i,j})}\to\sum_{i=1}^{n_l}{\int_{t_{l,i}^0}^{t_{l,i}}{F(\phi_{l,i})d\phi_{l,i}}}.
\end{equation}
In particular, the density of poles \ref{def_rinst} becomes a difference of sign functions with arguments $t-t_{l,i}$ and $t-t_{l,i}^0$. At this stage $n_l$ remains finite, it can be seen as a cut-off sent to infinity at the end of the computation. This is the point of view taken in \cite{Poghossian2010,*Fucito2011}. 

The support of the density $\brho\inst$ is rather complicated, and it is easier to work instead with the densities associated to the bounds $t_{l,i}^{0}$ and $t_{l,i}$ of the integration over $\phi_{l,i}$,
\begin{equation}
\rho_{t0}(t)=\sum_{l=1}^{N_c}\sum_{i=1}^{n_l}{\d(t-t_{l,i}^0)},\qquad \rho_{t}(t)=\sum_{l=1}^{N_c}\sum_{i=1}^{n_l}{\d(t-t_{l,i})}.
\end{equation}
These densities are normalized to $n=\sum_l{n_l}$, and exhibit a natural decomposition over the color index. In the NS limit the triple summation over the indices $l$ and $(i,j)\in Y_l$ becomes a single integration involving the difference of densities $\rho_t-\rho_{t0}$,
\begin{equation}\label{limit_NS_II}
\e_2\sum_{l=1}^{N_c}\sum_{(i,j)\in Y_l}{F(\phi_{l,i,j})}\to\int_\mathbb{R}{dt F^I(t)\left(\rho_{t}(t)-\rho_{t0}(t)\right)},
\end{equation}
where $F^I(t)$ is a primitive of $F(t)$. The constant of integration is irrelevant here since $\rho_t$ and $\rho_{t0}$ have the same norm. Upon integration by parts, the primitive $F^I(t)$ can be replaced by the original function $F(t)$, and the difference $\rho_{t0}-\rho_t$ by $\brho\inst$ since $\brho\inst'=\rho_{t0}-\rho_t$. The drawback of this method is the emergence of cumbersome boundary terms. We will not follow this path here, and simply work with the densities $\rho_t$ and $\rho_{t0}$.\\

We are now ready to take the NS limit of the instanton contribution \ref{sum_poles} to the partition function. As $\e_2\to0$, we introduce the effective action
\begin{equation}
\frac1{\e_2}\mathcal{S}\inst[\{\phi_I\}]=\dfrac{\e_2}{2}\sum_{I,J=1}^k{G(\phi_I-\phi_J)}+\sum_{I=1}^k\log\left(qQ_0(\phi_I)\right),
\end{equation}
such that the instanton partition function \ref{sum_poles} is approximated by 
\begin{equation}
 \CZ\inst(q,\e_1,\e_2,a_l,\mu_r)\simeq\sum_{\{Y_1,\cdots,Y_{N_c}\}}{e^{\frac1{\e_2}\mathcal{S}\inst}}.
\end{equation}
Kernel and potential of the action are simply derived from the original expression \ref{sum_poles},\footnote{The term coming from $\log D(t)$ has been symmetrized using (here $\phi_{IJ}=\phi_I-\phi_J$)
\begin{equation}
\sum_{I,J=1}^k{\log D(\phi_{IJ})}\simeq-\e_1\e_2\sum_{I,J=1}^k{\dfrac1{\phi_{IJ}(\phi_{IJ}+\e_1)}}=-\e_1\e_2\sum_{I,J=1}^k{\dfrac1{(\phi_{IJ}+\e_1)(\phi_{IJ}-\e_1)}}=\dfrac{\e_2}{2}\sum_{I,J=1}^k{G(\phi_{IJ})}.
\end{equation}
In the process, we neglected irrelevant constants coming from the initial subtraction of poles. Those constants are absorbed in the measure as the sum over Young tableaux is transformed into a functional integral \cite{Ferrari2012a}.}
\begin{equation}
Q_0(t)=\dfrac{M(t)}{A(t+\e_1)A(t)},\quad G(t)=(e^{\e_1\p_t}-e^{-\e_1\p_t})\dfrac1{t},
\end{equation}
where we introduced the convenient notation for the shift operator $e^{\a\p_t}g(t)=g(t+\a)$. Applying the procedure detailed above to take the NS limit, we obtain
\begin{equation}
\label{action_inst}
\mathcal{S}\inst=-\hf\int_\mathbb{R}{dt(\rho_t(t)-\rho_{t0}(t))\int_\mathbb{R}{ds(\rho_t(s)-\rho_{t0}(s))G^{II}(t-s)}}+\int_\mathbb{R}{dt(\rho_t(t)-\rho_{t0}(t))\int^t\log\left(qQ_0(t)\right)}
\end{equation}
with $G^{II}$ a double primitive of $G$, $\p^2G^{II}=G$. The function $G^{II}(t)$ is regular at $t=\pm\e_1$, however singularities may appear as we derive the equation of motion. The prescription for the regularization of poles at $t=s\pm\e_1$ is to shift $\e_1\to\e_1+i0$, as it corresponds to the case $\phi_I=\phi_J\pm\e_1$ in the original expression \ref{Zinst}. By definition, the instanton contribution to the free energy is 
\begin{equation}
\CF\inst=-\lim_{\e_2\to0}\e_1\e_2\log \CZ\inst
\end{equation}
In the saddle point approach, it is obtained as $\CF\inst\simeq-\e_1\mathcal{S}\inst^\ast$ where $\CS\inst^\ast$ denotes the evaluation of the action functional \ref{action_inst} for the critical density, i.e. the density that extremizes this action.

\paragraph{Correlation functions}
Using the localization technique, gauge theory correlators involving trace of powers of the adjoint scalar field $\Phi$ reduce to coupled one-dimensional integrals, just like the partition function \ref{Zinst} \cite{Poghossian2010}. They can be computed from the generating function
\begin{equation}\label{corr}
\la\tr e^{z\Phi}\ra=\sum_{l=1}^{N_c}e^{za_l}-(1-e^{z\e_1})(1-e^{z\e_2})\la\sum_{I=1}{e^{z\phi_I}}\ra\inst.
\end{equation}
On the RHS, the correlator is defined with the measure involved in the Nekrasov partition function,
\begin{equation}
\la\sum_I F(\phi_I)\ra\inst=\dfrac1{\CZ\inst}\sum_{k=0}^\infty{\dfrac{q^k}{k!}\left(\dfrac{\e_1+\e_2}{\e_1\e_2}\right)^k\int_\mathbb{R}{\prod_{I=1}^k{\dfrac{Q(\phi_I)d\phi_I}{2i\pi}}\prod_{\superp{I,J=1}{I\neq J}}^k{D(\phi_I-\phi_J)}\sum_{I=1}^{k}F(\phi_I)}}.
\end{equation}
In the NS limit, such correlators are approximated by
\begin{equation}\label{corr_NS}
\e_2\la\sum_I F(\phi_I)\ra\inst\simeq\int_{\mathbb{R}}{dtF^I(t)(\rho_t(t)-\rho_{t0}(t))},
\end{equation}
with $F^I$ a primitive of $F$, and the average performed with the critical density measure.\\

\noindent\textbf{Note: To avoid complicated expressions in the following sections, we drop the index $\e_1$ whenever the NS limit is considered, and simply write $\e=\e_1$.}

\subsection{Revealing the perturbative part}\label{Rev_pert}
In this subsection, we reorganize the densities in order to extract the perturbative part, allowing us to consider the full partition function in the subsequent sections. To simplify our considerations, we first take the case of pure SYM, i.e. $N_f=0$ and $M(t)=1$. In the saddle point approach, the Young tableaux columns (or the variables $t_{l,i}$) are seen as random objects achieving a statistical equilibrium. This equilibrium is obtained from the minimization of the effective action \ref{action_inst}. From this point of view, they play a role similar to the eigenvalues of a random matrix of size $n\times n$ with $n=\sum_ln_l$.\footnote{The method presented here is close but different from the one exposed in \cite{Klemm2008,*Sulkowski2009a,*Sulkowski2009} and based upon the transformation of sum over partitions with Plancherel measure into matrix integrals \cite{Eynard2008b}. The main difference lies in the definition of the eigenvalues since $a_l+\e_2(\l_i^{(l)}-i+1)$ is used instead of $t_{l,i}$ in \cite{Klemm2008}. It results in a model with a similar potential, but a different quadratic term.} This matrix decomposes over the color index into blocks of size $n_l$. The size of the blocks may also fluctuate, it is usually fixed by imposing filling fraction conditions in matrix models. The effective action $\CS\inst$ only depends on the difference $\rho_t-\rho_{t0}$. The density $\rho_t$ involves the random height $\l_i^{(l)}$ of the Young tableaux columns. On the other hand, the density $\rho_{t0}$ contains known parameters, the Coulomb branch vev $a_l$, as well as the ``random'' numbers of columns $n_l$. To single out the $n_l$-dependent part, we notice that the difference of $\rho_{t0}$ evaluated at $t$ and $t-\e$ is simply a sum of delta functions centered at $t=a_l$ and $t=a_l+\e n_l$. It leads us to split this density as $\rho_{t0}=\rho\ol-\rho\ct $ into a known part $\rho\ol$ and the random density $\rho\ct$ that satisfy respectively
\begin{equation}\label{def_rp}
(1-e^{-\e\p_t})\rho\ol(t)=\sum_{l=1}^{N_c}{\d(t-a_l)},\quad\text{and}\quad (1-e^{-\e\p_t})\rho\ct (t)=\sum_{l=1}^{N_c}{\d(t-a_l-\e n_l)}.
\end{equation}
There is an ambiguity in the definition of these densities which takes the form of an $\e$-periodic function. But this is actually irrelevant here as the densities will appear in the action \ref{action_inst} only through the well-defined difference $\rho(t)-\rho(t+\e)$. Moreover, this is not pertinent to our WKB approach with $\e$ infinitesimal for which the densities $\rho\ol$ and $\rho\ct $ are infinite series of distributions. The main issue is the treatment of diverging moments: they must be regularized using the principal value at infinity, and the infinite contributions from $\rho\ol$ and $\rho\ct $ integrals cancel each others. A more detailed discussion on these densities and infinities is provided in appendix \refOld{AppB}.\footnote{The decomposition of the instanton density \ref{def_rinst} $\brho\inst=\brho-\brho\ol$ into a perturbative density $\brho'\ol=-\rho\ol$ and $\brho'=-\rho$ has already been introduced in \cite{Ferrari2012a}, but without a clear interpretation. In the notations of \cite{Ferrari2012a}, they split $\rho=\rho_q+\rho_{Q_0}$ where $\rho_q$, $-\rho_{Q_0}$ and $\rho$ should be identified respectively with $\brho$, $\brho\ol$ and $\brho\inst$ here. The proper identification of $\brho\ol$ can be found in appendix \refOld{AppB}.}

The density $\rho=\rho_t+\rho\ct $ contains all the random variables. Formally, i.e. up to $a_l$-independent divergences, it is possible to choose for the expression of $\rho\ct $ a sum over delta functions centered at $t=a_l+(i-1)\e_1$ where $l$ runs over the number of colors and $i=n_l+1\cdots\infty$. Then, the density $\rho$ plays the role of $\rho_t$ for Young tableaux completed with an infinite number of columns of zero height,
\begin{equation}\label{rho}
\rho(t)=\sum_{l=1}^{N_c}{\sum_{i=1}^\infty{\d(t-a_l-(i-1)\e_1-\l_i^{(l)}\e_2)}},
\end{equation}
with $\l_i^{(l)}=0$ for $i>n_l$. The decomposition of $\rho$ into $\rho_t$ and $\rho\ct$ is related to the arctic circle phenomenon \cite{Eynard2008b}, and $\rho\ct$ describes the frozen eigenvalues below the arctic circle $n_l$.

However, we take here a different choice for $\rho\ct$ in order to keep norms finite. Our choice of $\rho\ct$ is obtained perturbatively in $\e$ in appendix \refOld{AppB}, it differs from the previous choice only by irrelevant ``boundary'' terms for the moments. The norm of the density $\rho\ol$ is also computed in appendix \refOld{AppB}, and the computation is similar for $\rho\ct$. Since the norm of $\rho_t$ is simply equal to $n$, we deduce
\begin{equation}\label{norms}
\int_\mathbb{R}{\rho(t)dt}=\int_\mathbb{R}{\rho\ol(t)dt}=-\dfrac1{\e}\sum_{l=1}^{N_c}{a_l}+\hf N_c.
\end{equation}
For a gauge group $SU(N_c)$, the sum over the Coulomb branch vevs $a_l$ is vanishing. Nevertheless, it is instructing to keep it explicit here. The first moments of $\rho\ol$ and $\rho$ are diverging, but the divergent terms compensate. Such boundary terms are independent of $a$, and only involve the parameter $\e$ and a cut-off. Thus, they do not contribute to the SW relations and can be discarded. We take
\begin{equation}\label{moments}
\int_\mathbb{R}{t\rho\ol(t)dt}=-\dfrac1{2\e}\sum_{l=1}^{N_c}{a_l(a_l-\e)},\quad\int_\mathbb{R}{t\rho(t)dt}=\dfrac1{2\e}\left(2\e_1\e_2k-\sum_{l=1}^{N_c}{a_l(a_l+2\e n_l)}\right)
\end{equation}
where we used the vanishing of the $a_l$ sum to simplify the second formula.\\

We now replace the difference of densities $\rho_t-\rho_{t0}$ in the action \ref{action_inst} with $\rho-\rho\ol$. The main result of this subsection comes from the observation that the cross-term between $\rho$ and $\rho\ol$ cancels the potential term involving $Q_0$ for $\rho$. To make it appear, we first fix the integration constant and define
\begin{equation}\label{def_GII}
G^{II}(t)=(e^{\e\p_t}-e^{-\e\p_t})\ t(\log t-1),
\end{equation}
such that both $G(t)$ and $G^{II}(t)$ are even functions. To evaluate the cross-term, we use a change of variable to transfer the shift operator from the kernel to the density $\rho\ol$, and then exploit the property \ref{def_rp}. Cancellation occurs with the potential provided we make a good choice for the integration constants, namely
\begin{equation}\label{ct}
\int_{\mathbb{R}^2}{dtds\rho(t)\rho\ol(s)G^{II}(t-s)}-\sum_{l=1}^{N_c}{\int_{\mathbb{R}}{dt\rho(t)(1+e^{\e\p_t})\int_{a_l}^t{\log(s-a_l)ds}}}=0.
\end{equation}
This choice was left arbitrary as long as the integration constants were the same for $\rho$ and $\rho\ol$. A similar property holds if we replace $\rho$ with $\rho\ol$ in the previous formula, it can be used to flip the sign of the quadratic term in $\rho\ol$. In the remaining expression of $\CS\inst$, densities are decoupled,
\begin{equation}\label{SSYM}
\CS\inst=\CS\SYM[\rho]-\CS\SYM[\rho\ol],\quad \text{with}\quad \CS\SYM[\rho]=-\hf\int_{\mathbb{R}^2}{dtds\rho(t)\rho(s)G^{II}(t-s)}+\log q\ \int_\mathbb{R}{dt\ (t+\e/2)\rho(t)}.
\end{equation}
We also fixed the integration constant for the primitive proportional to $\log q$ in the definition of $\CS\SYM$. Again, this choice is arbitrary provided that we make the same one for $\rho$ and $\rho\ol$.

It remains to interpret the action $\CS\SYM[\rho\ol]$. Discarding irrelevant boundary terms and divergences, the linear term proportional to $\log q$ reproduces the tree-level contribution to the free energy. To interpret the quadratic part, we need to introduce the function $\g_\e$ defined in the appendix and such that
\begin{equation}\label{rel_Gg}
\e G^{II}(t)=(e^{\e\p_t}-e^{-\e\p_t})(1-e^{\e\p_t})\g_\e(t).
\end{equation}
Once shift operators are transfered on the density by a change of variables, and eliminated through the formula \ref{def_rp}, the quadratic term in $\rho\ol$ reproduces exactly the one-loop vector contribution to the free energy, as found in \cite{Alday2009}. Thus, we have shown that the action $\CS\SYM[\rho\ol]$ supplies the perturbative (tree-level and one-loop) contribution to the free energy. Since by definition $\CS\inst$ provides the instanton contribution, the action $\CS\SYM[\rho]=\CS\inst+\CS\SYM[\rho\ol]$ generates the full free energy as $\CF\SYM=-\e\CS\SYM^\ast$. As $\rho$ is decoupled from $\rho\ol$, it is equivalent to extremize $\CS\inst$ or $\CS\SYM[\rho]$ to obtain the density $\rho$. Thus, the action $\CS\SYM[\rho]$ fully characterizes the partition function $\CZ\SYM$ in the NS limit. It is remarkable here that the very same action $\CS\SYM$ describes both perturbative and full partition functions. Knowing the definition \ref{def_rp} for $\rho\ol$, the action $\CS\SYM$ can be read from the perturbative free energy $\CF\ol$. This definition reduces to a sum of delta functions centered at $t=a_l$ for $\rho\ol'$ in the SW limit. Given its simplicity, it is natural to suppose that the definition \ref{def_rp} has a universal meaning. We conclude that the perturbative regime also determines the instanton contribution. It is of primary interest to figure out whether this is a general property of this set of $\mathcal{N}=2$ theories, and if it extends beyond the NS limit. Unfortunately, such questions are out of the scope of the present paper.

\paragraph{Turning on matter fields}
In the presence of fundamental matter fields, the mass polynomials $M(t)$ is responsible for a potential term in the action $\CS\inst$ given in \ref{action_inst}. Since no new cross-term appear, the decomposition of $\CS\inst$ into an action depending solely on $\rho$ and the same action with $\rho$ replaced by $\rho\ol$ still holds. However, this action also acquires a potential term,
\begin{equation}\label{action}
\CS\inst=\CS\full[\rho]-\CS\full[\rho\ol]\quad\text{with}\quad \CS\full[\rho]=-\hf\int_{\mathbb{R}^2}{dtds\rho(t)\rho(s)G^{II}(t-s)}+\int_\mathbb{R}{dt\rho(t)\int^t{\log\left(qM(t)\right)}}.
\end{equation}
For a good choice of the integration constant, this new linear term in $\CS\full[\rho\ol]$ generates the massive contribution to the perturbative free energy at one loop. Consequently, the full free energy, including instanton and perturbative terms, is again given by the action depending only on $\rho$ as $\CF\full=-\e\CS\full[\rho]$.

In \cite{Nekrasov2003a}, a trick was employed to re-introduce the mass term from the pure SYM action using a shift of the field by an appropriate mass term. This trick generalizes to the NS limit using a density $\rho\mass$ solving\footnote{To make contact with the next subsection, we may define a function $f\mass$ associated to $\rho\mass$ through a relation similar to \ref{fri} below, i.e.
\begin{equation}
(1+e^{\e\p_t})f\mass''(t)=\sum_{r=1}^{N_f}{\d(t-\m_r)},
\end{equation}
In the SW limit $\e\to0$, we recover the definition used in \cite{Nekrasov2003a} (equ 7.3).}
\begin{equation}
(e^{\e\p_t}-e^{-\e\p_t})\rho\mass(t)=-\sum_{r=1}^{N_f}{\d(t-\m_r)}.
\end{equation}
The action $\CS\full[\rho]$ can be written as the difference of the SYM actions $\CS\SYM[\rho+\rho\mass]-\CS\SYM[\rho\mass]$. In this expression, the cross term between $\rho$ and $\rho\mass$ generates the massive potential for $\rho$ in the same way $\rho\ol$ has generated the potential term depending on the Coulomb branch vevs in the previous calculation. The linear term, and possible divergences, are eliminated by subtracting $\CS\SYM[\rho\mass]$. At several occasions we shall restrict ourselves to pure SYM, and this trick can be employed to recover the massive case.

\subsection{Effective action for the shape function}
Previously, the Young tableaux dependence has been encoded in the instanton density $\brho\inst$, later traded for the more appropriate function $\rho$. In the work of Nekrasov and Okounkov \cite{Nekrasov2003a}, a different collective field is used. There, a shape function $f_Y$ is associated to the Young tableau $Y=(\l_1\geq\l_2\geq\cdots\geq\l_n)$ as\footnote{Our notation differs from \cite{Nekrasov2003a}, $-2f_\text{here}=f_\text{[NO]}$. Note also that the summation can be extended to $i=1\cdots\infty$ if we set $\l_{i>n}=0$.}
\begin{equation}\label{density_profile}
f_Y(t|\e_1,\e_2)=-\hf|t|-\hf(1-e^{-\e_1\p_t})\sum_{i=1}^n{\Big[|t-(i-1)\e_1-\l_i\e_2|-|t-(i-1)\e_1|\Big]}.
\end{equation}
The function $f_Y$ describes the profile of the tableau $Y$ rotated such that it occupies the angular sector $[\pi/4,3\pi/4]$. Following \cite{Nekrasov2003a}, we assume that the Coulomb branch vevs are real, ordered as $a_1<a_2<\cdots<a_l$, and separated by a distance greater than the Young tableau size $\e_2\l_1^{(l)}\times\e_1 n_l$ such that the support of the functions $f_{Y_l}''(t-a_l)$, centered at $t=a_l$, do not intersect. Cases where these conditions are not satisfied may be obtained from analytical continuation. We then define the total shape function $f$ as the sum of Young tableaux profiles,
\begin{equation}\label{shape}
f(t)=\sum_{l=1}^{N_c}{f_{Y_l}(t-a_l)}.
\end{equation}
In the NS limit, $f(t)$ is treated as a collective field determined from the minimization of an effective action.

Our approach to the problem provides a bridge between the formalisms used in \cite{Nekrasov2003a} and \cite{Poghossian2010}, and we will work indifferently with shape functions $f$ or densities $\rho$. To relate these two quantities, we note that as $\e_2\to0$ shape function and instanton density obey
\begin{equation}\label{rel_fr}
f'(t)=f'\ol(t)+(1-e^{-\e_1\p_t})\brho\inst(t),\quad\text{with}\quad f\ol(t)=-\hf\sum_{l=1}^{N_c}{|t-a_l|}.
\end{equation}
The functions $f\ol$ and $\rho\ol$ trivially satisfy the relation \ref{fri} given in the introduction. To show that it also true for $f$ and $\rho$, it suffices to differentiate \ref{rel_fr} and use the fact that $\brho\inst'=\rho\ol-\rho$. This formula can be used to replace the shape function by differences of the density $\rho$. But it can also be inverted, at least perturbatively in $\e$, to trade the density for the shape function.

The effective action over shape functions is derived from the expression \ref{SSYM} of $\CS\SYM[\rho]$. To do so, the kernel $G^{II}$ is written as a difference of $\tg_\e$ functions using \ref{G_tge}. Shift operators are transposed to the densities with a change of integration variable, and produces shape functions through \ref{fri}. We end up with the action \ref{CSSYM_f} given in the introduction. As already mentioned, matter can be reintroduced by adding a mass term to the function $f$, and subtracting the pure mass action. The norm and first moment of the shape function follow from the primary definition \ref{shape} of $f$,
\begin{equation}\label{prop_f}
\int_\mathbb{R}{f''(t)dt}=-N_c,\qquad \int_\mathbb{R}{tf''(t)}=-\sum_{l=1}^{N_c}{a_l}.
\end{equation}
These formulas are in agreement with the relation \ref{fri} provided we take into account the boundary terms (see appendix \refOld{AppB}).

Sending $\e\to0$, we recover the results obtained in the study of the SW limit. In particular, the action $\CS\SYM[f'']$ given in \ref{CSSYM_f} reduces to the effective action used in \cite{Nekrasov2003a}. The solution $f_0$ to the equation of motion is called the ``limit shape''. It was found in \cite{Nekrasov2003a} and will be re-derived in the next subsection. Upon integration of \ref{rel_fr}, we also recover the relation between $f_0$ and the instanton density, $f_0=f\ol+\e_1\brho\inst$ (equ 4.37 in \cite{Nekrasov2003a}). From \ref{fri}, we deduce $-f_0''(t)=\e\rho_0'(t)$ where the subscript $0$ denotes the first order in the $\e$-expansion of the density $\rho$. The limit shape $f_0$ exhibits a finite support $\G$ being the union of $N_c$ intervals $\G_l=[\a_l^-,\a_l^+]$ such that $a_l\in\G_l$. On each of these intervals, $-f_0''$ plays the role of a density and vanishes at the endpoints $\a_l^\pm$. It is positive if $f_0$ is concave, which may not be true everywhere. However, the positivity of the density is not required in the context of SW theory \cite{Chekhov2002}. The normalization of $f_0''$ is also given by \ref{prop_f} since the RHS is independent of $\e$.

In the NS limit, the limit shape acquires $\e$-corrections that will be obtained below by solving perturbatively the equation of motion associated to the action \ref{CSSYM_f}. We shall show that each term of the $\e$-expansion $f''=\sum_{n=0}^\infty{\e^nf_n''}$ share the same support $\G$ as the first order term $f_0''$. Due to the formula \ref{fri}, it is also true of the terms $\rho'_n$ in the expansion of $\rho'$. Moreover, both $\rho$ and $f''$ have a natural decomposition over the color index, and each component is vanishing on all but one interval $\G_l$. We deduce from \ref{prop_f} the ``filling fraction'' conditions
\begin{equation}\label{fillings}
\int_{\G_l}{\rho(t)dt}=-\dfrac{1}{\e}(a_l-\e/2),\quad \int_{\G_l}{f''(t)dt}=-1,\quad  \int_{\G_l}{tf''(t)}=-a_l.
\end{equation}

\section{Equation of motion and SW relations}
\subsection{Definition of the resolvents}
In the study of matrix models planar limit, it is usual to introduce a resolvent associated to the density of eigenvalues. This quantity is known to have a branch cut on the support of the density, along which it has a discontinuity of $-2i\pi$ times the density. For a large class of potentials, the resolvent satisfies an algebraic equation that defines the spectral curve. In the context of \bens s, this algebraic relation becomes a first order non-linear differential equation, the Riccati equation. As reviewed in the subsection \refOld{sec_bens} below, such equations can be put in the form of a Schr\"odinger problem. Following the same approach, we introduce here the wave function $\hpsi$ such that $\p\log\hpsi(z)$ is the resolvent associated to $\rho$,
\begin{equation}
\log\hpsi(z)=\int_{\mathbb{R}}{\rho(t)\log(z-t)dt}.
\end{equation}
Since we can equivalently work with the shape function, we also define a wave function $\o(z)$ such that $\p\log\o(z)$ is the resolvent associated to $f''$,
\begin{equation}\label{def_o}
\log\o(z)=\int_{\mathbb{R}}{f''(t)\log(z-t)dt}.
\end{equation}
These two quantities coincide with the ones employed in \cite{Poghossian2010}. As a consequence of \ref{fri}, the wave functions $\hpsi$ and $\o$ are related through
\begin{equation}\label{ratio_psi}
\o(z)=\dfrac{\hpsi(z-\e)}{\hpsi(z)}.
\end{equation}
Placing the branch cut of the logarithm on $\mathbb{R}^-$, the function $\log\hpsi(z)$ is discontinuous on the interval $]-\infty,\a_{N_c}^+]$, and values above and below the cut read
\begin{equation}\label{prop_log_res}
\log\hpsi(t\pm i0)=\int_\mathbb{R}{\rho(s)\log|t-s|}\pm i\pi \int_t^\infty{\rho(s)ds}.
\end{equation}
Deriving this relation with respect to $t$, we recover the standard Feynman poles prescription for the resolvent, provided we identify the principal value with the derivative of the weakly singular integral with kernel $\log|t-s|$. The function $\hpsi$ has the same branch cut as its logarithm. On the other hand, a simplification occurs for $\o$ which has a branch cut only on the support $\G$ of $f''$. The reason is that the discontinuity of $\log\o$ is constant outside of the support $\G$, since its derivative is the resolvent of a density supported on $\G$. Constants on each side of an interval $\G_l$ differ by $2i\pi$ times the integral of the density over this interval, which gives $-1$ in the case of $f''$ (see equ \ref{fillings}). There is no discontinuity of $\log\o$ on the interval $[\a_{N_c}^+,\infty[$, and these constants must be equal to a multiple of $2i\pi$. Upon exponentiation, we conclude that $\o$ has no discontinuity outside of the support $\G$.

As noticed in \cite{Poghossian2010}, the wave function $\o$ is related to correlators with insertion of the adjoint scalar field $\Phi$. Indeed, in the NS limit the correlator \ref{corr} can be computed using the formula \ref{corr_NS}. Replacing the difference of densities by $\rho-\rho\ol$, the perturbative contribution cancels the first term in \ref{corr}, and we end up with
\begin{equation}\label{corr_Phi}
\la\tr e^{z\Phi}\ra=-\int_\mathbb{R}{dte^{zt}f''(t)}.
\end{equation}
Expanding in $z$, we deduce the moments
\begin{equation}\label{def_amp}
\la\tr 1\ra=N_c,\quad \la\tr \Phi\ra=\sum_{l=1}^{N_c}{a_l}=0,\quad u_s=\la\tr\Phi^s\ra=-\int_\mathbb{R}{dt t^s f''(t)},
\end{equation}
where $u_s$ with $s=2\cdots N_c$ parameterize the gauge theory vacua. Similar relations were obtained in \cite{Dijkgraaf2002a} for the matrix model describing $\mathcal{N}=2$ SYM on $\mathbb{R}^4$. By identification of the terms in its $z$-expansion, we relate $\o$ to the amplitude
\begin{equation}
\o(z)^{-1}=\exp\la\tr\log(z-\Phi)\ra\simeq \la\det(z-\Phi)\ra.
\end{equation}
In this formalism, the Matone relation \cite{Matone1995,*Sonnenschein1996,*Eguchi1996,*Flume2004,*Matone2002} is easily derived from the action \ref{CSSYM_f},
\begin{equation}
q\dfrac{\p\CF\full}{\p q}=\hf\la\tr\Phi^2\ra.
\end{equation}

\subsection{Equation of motion}
The critical density $\rho$ minimizing the action \ref{action} solves the equation of motion\footnote{Working with the shape function $f$, we should instead consider
\begin{equation}
(1-e^{\e\p_t})\p_t\dfrac{\d\CS\full[f'']}{\d f''(t)}=0.
\end{equation}}
\begin{equation}\label{eom}
\p_t\dfrac{\d\CS\full}{\d\rho(t)}=0\implies \int_\mathbb{R}{ds\rho(s)G^I(t-s)}=\log\left(qM(t)\right).
\end{equation}
In terms of the wave functions $\hpsi$ or $\o$ defined above, the equation of motion reads,
\begin{equation}\label{eom2}
 qM(t)\dfrac{\hpsi(t-\e)}{\hpsi(t+\e)}=1\quad\Leftrightarrow\quad qM(t)\o(t)\o(t+\e)=1.
\end{equation}
Those equations were previously established in \cite{Poghossian2010}, they are valid for finite $\e$ and only holds at the discrete set of points $t\in\cup\{t_{l,i}\}$. In the case of an infinitesimal $\e$, the support of the density condenses into the union of intervals $\G$. In this limit, we need to introduce a regularization of the singularities at coincident instantons $\phi_I=\phi_J$. It results in the presence of an additional cut-off term in the effective action. Another cut-off contribution may also come from the replacement of the sum over Young tableaux with a functional integral over the density $\rho$. Such cut-off terms were first observed on matrix models by Dyson \cite{Dyson1962}. More recently, they have been considered in the context of the Dijkgraaf-Vafa \bens\ in \cite{Bourgine2012}.\footnote{See also \cite{Wiegmann2005} for a discussion in the framework of the 2D Coulomb gas.} In this model, the terms coming from the regularization of kernel and measure cancel in the SW limit where $\b=1$, and we recover the standard Hermitian matrix model. When $\b\neq 1$, i.e. in the NS limit, they should be taken into account as they reproduce the derivative term of the loop equations. More details on this phenomenon, and explicit formulas, will be provided in section \refOld{sec_bens} below.

Our strategy will be as follows. We first concentrate on the SW limit, and suppose that there are no cut-off terms. We will derive an equation of motion valid on the whole complex plane, and recover the SW curve which validates our assumption. We then turn to the NS limit, and consider the equation of motion \ref{BaxterTQ} derived from AGT conjecture in the next section. We deduce the expression of the cut-off term to be included in the effective action. This expression is further motivated by rewriting the kernel part of the action \ref{CSSYM_f}.

\subsubsection{Seiberg-Witten limit}
In the SW limit, we expand the kernel $G^I(t)\simeq 2\e/t$ in \ref{eom} and the integral is considered as a principal value. Actually, it is more convenient to work with the shape function $f_0''=-\e\rho_0'$, and focus on the equation of motion
\begin{equation}
2\int_\mathbb{R}{f_0''(t)\log|t-s|ds}+\log\left(qM_0(t)\right)=0,\quad t\in\G.
\end{equation}
A possible $\e$-dependence of the mass polynomial may appear in the presence of hypermultiplets in the anti-fundamental, and we denoted $M_0$ its SW limit. The wave function $\o_0$ obeys the property \ref{prop_log_res} with $\rho$ replaced by $f_0''$. Thus, the integral in the previous equation is simply the sum of the values of $\log\o_0$ above and below the branch cut. Exponentiating the equation, we deduce
\begin{equation}
qM_0(t)\o_0(t+i0)\o_0(t-i0)=1,\quad t\in\G.
\end{equation}
It implies that the quantity
\begin{equation}\label{def_P0}
P_0(z)=\dfrac1{\o_0(z)}+qM_0(z)\o_0(z)
\end{equation}
has no discontinuity on $\G$. The asymptotic at $z\to\infty$ can be derived from the equations \ref{prop_f},
\begin{equation}\label{asympt_o}
\log\o(z)\sim -N_c\log z+\dfrac1z\sum_{l=1}^{N_c}{a_l}+O\left(\dfrac1{z^2}\right)\implies P_0(z)\sim z^{N_c}+q z^{N_f-N_c}.
\end{equation}
Since there is no reason to expect any other singularity for $P_0(z)$, it should be a polynomial of degree $N_c$ when $N_f\leq 2N_c$. In the following equations, we assume that $P_0(z)$ is a monic polynomial, and consequently when $N_f=2N_c$, $P_0(z)$ should be replaced by $(1+q)P_0(z)$. The relation \ref{def_P0} between $\o_0$ and $z$ defines a double cover of a sphere with $N_f$ punctures at $z=\mu_r$. This curve is identified with the SW curve of the gauge theory on $\mathbb{R}^4$ \cite{Nekrasov2003a}. This result justifies the assumption that no cut-off term is present in the effective action for the SW limit. The function $\o_0$ is multivalued and should be defined on a hyperelliptic surface of genus $N_c$ with singularities at the poles of $M_0(z)$. The two solutions of the quadratic equation \ref{def_P0} give the value of $\o_0$ on the two sheets covering the punctured sphere,
\begin{equation}\label{sol_o0}
\o_0^\pm(z)=\dfrac{P_0(z)\pm\sqrt{P_0(z)^2-4qM_0(z)}}{2qM_0(z)}.
\end{equation}
The physical, or principal, sheet is by definition the sheet where $\o_0$ satisfies the asymptotic expansion \ref{asympt_o}. The value of $\o_0$ on this sheet is given by $\o_0^-$. The $2N_c$ branch points $\a_l^{\pm}$ are such that
\begin{equation}\label{poly_bp}
\prod_{l=1}^{N_c}(z-\a_l^+)(z-\a_l^-)=P_0(z)^2-4qM_0(z),
\end{equation}
which is indeed the equation (4.50) of \cite{Nekrasov2003a}. In the superconformal case, the LHS must be multiplied by $(1-q)^2$ and $P_0$ replaced by $(1+q)P_0$.

\subsubsection{Next orders in $\e$}
To go beyond the SW limit, we assume that the quantity $P$ appearing in 
\begin{equation}\label{saddle_o}
qM(z)\o(z+\e)\o(z)-P(z)\o(z+\e)+1=0
\end{equation}
has no branch cut over $\G$. This is motivated by the study at finite $\e$ performed in \cite{Poghossian2010}. However, a minor difference appears here in the sign of $q$. This sign flip is necessary to reproduce the correct SW curve as $\e\to0$. It is also in agreement with the AGT conjecture, as shown in the next section. The arguments employed on $P_0$ can be repeated to show that $P(z)$ must be a monic polynomial of degree $N_c$, and has to be replaced by $(1+q)P(z)$ when $N_f=2N_c$. Trading $\o$ for the ratio \ref{ratio_psi} of $\hpsi$ functions, the equation \ref{saddle_o} produces a difference equation analogue to the Baxter TQ relation,
\begin{equation}\label{BaxterTQ}
qM(z)\psi(z-\e)-P(z)\psi(z)+\psi(z+\e)=0.
\end{equation}
We now relate the condition of vanishing discontinuity for $P(z)$ to the equation of motion. Our claim is that this is implied by the equation of motion \ref{eom}, provided we correct it to account for the cutoff term in the collective action. The requirement $\Disc P=0$ on $\G$ is equivalent to the condition,
\begin{equation}
q^2M^2\o^+\o^-\o_\e^+\o_\e^-F=1\quad\text{with}\quad F=\dfrac{(\o^+-\o^-)^2}{\o^+\o^-}\dfrac{\o_\e^+\o_\e^-}{(\o_\e^+-\o_\e^-)^2},
\end{equation}
where $\o_\e$ is a shortcut notation for the function with shifted argument $\o(t+\e)$, and the superscript $\pm$ denotes the position with respect to the branch cut, $\o^\pm=\o(t\pm i0)$. The principal value regularization of the equation of motion derived from the action $\CS\full$ produces the same identity with $F=1$. The multiplicative alteration of the equation of motion is the sign of an additive correction to the action. Since a cutoff term arises from the regularization of kernel and measure, it must be independent of $q$ or $M$, which is indeed the case of the function $F$. In addition, cut-off terms usually involve the density by itself, and not the principal value of the resolvent. It implies that $F$ must depend only on the ratio $\o^+/\o^-$, possibly shifted by $\e$, and cannot depend on the product $\o_+\o_-$. The expression we found for $F$ again satisfies this property. Put it otherwise, the cut-off terms we add here cannot be replaced by a modification of kernel or potential. 

The function $F$ writes in terms of the limit shape function,
\begin{equation}
F(t)=\dfrac{\sin^2\pi f'(t)}{\sin^2\pi f'(t+\e)}\implies \log F(t)=2(1-e^{\e\p_t})\log|\sin(\pi f'(t))|.
\end{equation}
We notice that $F$ is real. We also verify that as $\e\to0$, $F=O(\e)$ gives no contribution in the SW limit. In general, at the order $O(\e^n)$ where the equation determines $f_n''$, the cut-off term involves only shape functions of a smaller order $f_{k<n}''$. The corrected saddle point equation reads
\begin{equation}\label{correct_eom}
(1+e^{\e\p_t})\int_\mathbb{R}{f''(s)\log|t-s|}+(1-e^{\e\p_t})\log|\sin(\pi f'(t))|+\log qM(t)=0.
\end{equation}
To reproduce the amended equation of motion, the collective field action $\CS\full$ should be modified as
\begin{equation}\label{CS_cutoff}
\CS\full[f'']\to\CS\full[f'']+\int_\mathbb{R}{l^I(f'(t))dt},\qquad l^I(x)'=\log|\sin(\pi x)|.
\end{equation}
To understand the expression of the cut-off term, we need to investigate the quadratic part of the action. We notice that unlike the kernel $G^{II}$ associated to the density $\rho$, the kernel $\tg_\e$ is not an even function, but satisfies
\begin{equation}
\tg_\e(-x)=\tg_\e(x)-\e^2 l^I(x/\e).
\end{equation}
It allows to rewrite the quadratic term of the action as
\begin{equation}
-\dfrac1{\e}\int_{\mathbb{R}^2}{dtdsf''(t)f''(s)\tg_\e(t-s)}= -\dfrac1{\e}\int_{\mathbb{R}^2}{dtdsf''(t)f''(s)\tg_\e(|t-s|)}-\e\int_{t<s}{dtdsf''(t)f''(s)l^I((t-s)/\e)}.
\end{equation}
The second term in the RHS is weakly divergent for $t=s$, its regularization produces the cut-off term in the action \ref{CS_cutoff} with a cut-off length $\sim \e f'(t)$.

Eventually, our claims concerning the support of densities are justified by the study of the equation \ref{saddle_o}. This equation has to be understood as a series expansion in $\e$ for the quantities $\o$, $P$ and $M$. At the first order, we recover the two solutions \ref{sol_o0} for $\o_0$. At the next to leading order, it provides the correction
\begin{equation}
\o_1=-\dfrac{qM_1\o_0^2+qM_0\o_0\o_0'-P_1\o_0-P_0\o_0'}{2qM_0\o_0-P_0},
\end{equation}
where the subscripts denote the order in the $\e$-expansion. This function $\o_1$ shares the same branch cut as $\o_0$ and its derivative. It is actually easy to see that at the order $O(\e^n)$, the equation is always linear in $\o_n$ and involves no derivative of this quantity. Thus, $\o_n$ is simply determined by the lower order functions $\o_{k<n}$ and their derivatives, together with $P_k$ and $M_k$. We conclude by recursion that all functions $\o_n$ have the same branch cut as $\o_0$. They may however involve branch points of higher order, but always of the square root type, i.e. $(z-\a_l^\pm)^{r/2}$ with $r\in\mathbb{Z}$. Consequently, all correction orders $f_n''$ to the limit shape, and $\rho_n'$ to the density $\rho'$ after inverting \ref{fri}, share the same support $\G$. As for the \bens, this may not be true anymore when we consider the resummed series at finite $\e$ \cite{Eynard2008a,*Chekhov2009,*Chekhov2010}.

\subsubsection{Discussion}
So far a rigorous derivation of the action cut-off term has not been provided. It would indeed be more satisfying to first determine the correct effective action, including the cut-off term, through a more careful treatment of divergences and measure. The Baxter TQ relation \ref{BaxterTQ} would then follow from the action's minimization, and the next section provide a proof of AGT correspondence at $\e_2=0$. Another possibility to derive the relation \ref{BaxterTQ} is the use of loop equation techniques. This path was followed in \cite{Bourgine2012} for the study of the \bens. Loop equations are specific Schwinger-Dyson equations obtained by exploiting the invariance of the integration measure. Such equations can be established for the Nekrasov partition function from its expression \ref{Zinst} as contour integrals. The main difficulty consists in writing a closed and amenable set of equations in the NS limit. We hope to address this issue in a near future.

On the CFT side, loop equations are known to be related to the invariance of the Liouville correlator, or the \bens\ partition function, under the generators of a Virasoro algebra. On the other hand, Virasoro generators seems to act on the Young tableaux summations involved in the Nekrasov partition function by adding or subtracting boxes, as suggests the study performed at $\b=1$ in \cite{Kanno2012}. It would be interesting to understand how this interpretation may show up in our formalism. Finally, another major motivation to develop the loop equation technique is the possibility to access to $\e_2$ corrections. Like for the matrix models, it must be possible to derive a tower of loop equations that can be solved recursively, possibly employing a generalization of the Eynard-Orantin topological recursion.

\subsection{Deformed Seiberg-Witten relations}
We now proceed to the derivation of $\e$-deformed SW relations. We follow the approach used by Chekhov and Mironov on the Hermitian matrix model \cite{Chekhov2002}. This derivation was later extended to the Dijkgraaf-Vafa \bens\ in \cite{Bourgine2012}. The major difference is the presence of a cut-off term in the effective action close to the one obtained in \ref{CS_cutoff}. This term results in poles contribution at the branch points which modify the SW relations. From this point of view, the treatment presented here is very similar to the case of \bens s.

The dependence of the collective action in the Coulomb branch vevs $a_l$ is encoded in the density, and the derivative of the action with respect to $a_l$ is simply
\begin{equation}\label{dSdal}
\dfrac{\p\CS\full}{\p a_l}=\int_\mathbb{R}{dt\dfrac{\p\rho}{\p a_l}(t)C(t)}\quad\text{where}\quad C(t)=\dfrac{\d\CS\full}{\d\rho(t)}.
\end{equation}
As a consequence of the equation of motion \ref{eom}, $C(t)$ is a constant $c_l$ on each interval $\G_l$. In the context of (deformed) SW theory, these constants should be allowed to be different \cite{Chekhov2002}.\footnote{In the SW limit, the constants $c_l$ are related to the Lagrange multipliers $\xi_l$ introduced in \cite{Nekrasov2003a}. This can be seen after writing $C(t)$ with the help of the limit shape $f_0''(t)$. Taking here $M(t)=1$ for simplicity, the condition 
\begin{equation}
C(t)=2\int_\mathbb{R}{dsf_0''(s)(t-s)\left(\log\dfrac{|t-s|}{\L_\text{[NO]}}-1\right)}=c_l,\quad\text{for}\quad t\in\G_l,
\end{equation}
is the analogue of the equation (4.40) in \cite{Nekrasov2003a}.} The remaining integrals can be evaluated by differentiating the relation \ref{fillings}, and the partial derivative of $\CS\full$ with respect to $a_l$ is simply equal to $-c_l/\e$. The constants $c_l$, or more precisely their differences, can be obtained as cycle integrals of a  $1$-form over the SW curve. To do so, we need to define a function $\tC(z)$ on this curve, i.e. a bi-valued function on $\mathbb{C}$ with a branch cut on $\G$. We require that the sum of the values above and below the branch cut $\G_l$ is equal to $2c_l$. We also introduce the cycles $A_l$ that circle the branch cuts $\G_l$ in the physical sheet, and the dual cycles $B_l$ that satisfy $A_k\cap B_l=\d_{k,l}$ where indices range from $2$ to $N_c$. The cycles $B_l$ relate the two sheets by passing through the first and $l$th branch cuts. The difference of constants $c_l$ can be expressed as a cycle integral over $B_{l+1}-B_l$ which is passing through the $l$th and $(l+1)$th branch cuts,
\begin{equation}
\dfrac{\p\CS\full}{\p a_l}-\dfrac{\p\CS\full}{\p a_{l+1}}=\dfrac1{2\e}\oint_{B_{l+1}-B_l}{\p \tC(z)dz},\quad l=1\cdots N_c-1,
\end{equation}
with the trivial cycle $B_1$. In this expression, $\p\tC$ can be shifted by any meromorphic function $v(z)$ on $\mathbb{C}$ provided it has no poles inside the integration contour, i.e. at the branch points. This property, which also holds for $A_l$-cycle integrals, can be used to simplify the definition of the SW differential. We will set $2dS=d\tC+v(z)dz$ with $v(z)$ chosen appropriately. Taking into account that for an $SU(N)$ gauge group, the sum of $a_l$-partial derivatives of the free energy is vanishing, we deduce the SW relation
\begin{equation}
\dfrac{\p\CF\full}{\p a_l}=\oint_{B_l}{dS}.
\end{equation}

We now determine $\tC(z)$. Since we are only interested in the differences $c_l-c_{l+1}$, it is simpler to directly look for $\p\tC(z)$. It is also more convenient to work here with the limit shape $f$ instead of the density $\rho$. The expression of $C(t)$ is obtained by evaluating the variation of the action with respect to $\d\rho(t)$, and then replacing the remaining densities by shape functions using \ref{fri}. It suggests to use
\begin{equation}
\p\tilde{C}(z)=(1+e^{\e\p})\log\o(z)+(1-e^{\e\p})L(z)+\log qM(z),
\end{equation}
with a function $L(z)$ defined on the SW curve, and satisfying for $t\in\G$ the relation
\begin{equation}
L(t+i0)+L(t-i0)=2\log|\sin(\pi f'(t))|.
\end{equation}
At the origin of the function $L(z)$ is the presence of the cut-off term in the action \ref{CS_cutoff}. A similar function, also denoted $L$, appeared in the treatment of the DV \bens\ \cite{Bourgine2012}. It can be expressed with the help of another function $b(z)$, also defined on the SW curve, and with a vanishing real part on $\G$,
\begin{equation}
L(z)=\log b(z),\quad\text{with}\quad b(t\pm i0)=\pm i\sin\pi f'(t)\quad\text{for}\quad t\in\G.
\end{equation}
The function $b(z)$ is a solution to the quadratic equation
\begin{equation}
b^2=\dfrac14\dfrac{(\o^+-\o^-)^2}{\o^+\o^-},
\end{equation}
where the functions $\o^\pm(z)$ are the values of $\o$ on the two sheets. They can be obtained recursively using the equation \ref{saddle_o} and starting respectively with the functions $\o_0^\pm(z)$ found in \ref{sol_o0}. We observe that the discontinuity of $L$ over $\G_l$ is a constant, and it gives no contribution to $\p\tC$. However, to suppress the function $L$ in the definition of the SW differential $dS$, we further have to show that it has no pole at the branch points. It turns out that such poles exist at the first order,
\begin{equation}\label{corr_1st}
(1-e^{\e\p})L(z)\sim\dfrac{\e}2\p\log M_0(z)-\dfrac{\e}{2}\sum_{l,\pm}\dfrac1{z-\a_l^\pm}.
\end{equation}
This term is responsible for shifts of $-\e/2$ for the $A_l$-and $B_l$-cycle integrals. It is possible to show that there are no pole contributions at higher orders.\footnote{The proof goes as follows. We denote $\CF[R]$ the ring of functions of the form $A+BR$ where $A$ and $B$ are meromorphic and $R$ is the square root of the polynomial given in \ref{poly_bp}, with branch points at $\a_l^\pm$. This set is stable under multiplication, ratio, and differentiation. We easily check that $\o_0$, and by recursion all $\o_n$ belong to this set.  It is also true of $b^2$ at any order. Expanding the logarithm, we find that $L=\log b_0+\CF[R]$ where $b_0$ is the first order in the $\e$-expansion of the function $b$, and $\CF[R]$ indicates the presence of a function in this set. We deduce
\begin{equation}
(1-e^{\e\p})L(z)=-\e\p\log b_0(z)+\p\CF[R].
\end{equation}
The first term generates the $O(\e)$ correction presented in \ref{corr_1st}. The second term designates the derivative of a function in $\CF[R]$. It has no residues at the branch points and gives no contribution to cycle integrals.} To simplify later comparison in the context of AGT conjecture, we neglect here the function $L$ in the definition of the SW differential, and take
\begin{equation}\label{def_SW}
dS=\hf \log(\o(z)\o(z+\e))dz.
\end{equation}
The counterpart is the presence of the $\e$-shifts in the relations, as seen in the first relation \ref{SW_B}. Upon integration by parts, we recover in the SW limit the usual $1$-form $dS=-zd\o/\o$, and \ref{SW_B} reproduces the standard SW relations.

Once the differential form has been determined, it remains to verify the $A_l$-cycle relation. Since $\p\log\o$ is the resolvent associated to the density $f''(t)$, and given the properties \ref{fillings}, we deduce the second expression in \ref{SW_B}. The $\e$-correction was first observed in \cite{Nishinaka2011} in the context of the Dijkgraaf-Vafa matrix model. As for the first relation, it is the consequence of poles at the branch points, and can be absorbed by including the function $L$ within the definition of the SW differential.

In the next section, we investigate a transformation of the SW differential into the equivalent $1$-form for the DV \bens. It will be simpler there to consider the wave function $\hpsi$ instead of $\o$. We claim that upon a shift of integration variables the differential \ref{def_SW} is equivalent to 
\begin{equation}\label{def_SW2}
dS=-\e\p\log\hpsi(z)dz.
\end{equation}
It is easy to check using \ref{fillings} that it gives the same result for the $A_l$-integral.

\section{AGT relation in the NS limit}
In order to compare our results with the DV \bens\ \cite{Dijkgraaf2009}, we restrict ourselves to the gauge group $SU(2)$. Our aim is to relate the density of eigenvalues of this model, or more precisely the associated wave function $\psi$, to the corresponding quantity on the gauge theory side. Higher rank gauge groups can be considered at the price of introducing a $\b$-deformed matrix chain \cite{Itoyama2009}. These models involve several eigenvalue densities. Since only a single density is needed for the Nekrasov instanton partition function, the matching becomes more involved. On the other hand, the quantum change of variable presented here easily generalizes to such more complicated cases.

For simplicity, we will only treat the case of four flavors: two in the fundamental representation of the gauge group with masses $\mu_\pm$, and two in the anti-fundamental with masses $\tilde{\mu}_\pm$. Models with a smaller number of flavors can be obtained by taking a suitable limit \cite{Gaiotto2009b,*Marshakov2009,*Itoyama2010}. Since the equality $N_f=2N_c$ is satisfied, the model is superconformal. It is sometimes referred to as superconformal QCD, or SCQCD. The mass and gauge polynomials entering into the expression \ref{Zinst} of the instanton contribution to the partition function are respectively of degree four and two,
\begin{equation}\label{def_M}
M(t)=(t-\mu_+)(t-\mu_-)(t-\e+\tilde{\mu}_+)(t-\e+\tilde{\mu}_-),\qquad A(t)=t^2-a^2.
\end{equation}

\subsection{Dijkgraaf-Vafa \bens\ and Liouville theory}\label{sec_bens}
To introduce the DV \bens, we start from the Liouville correlator of four vertex operators inserted at the points $0,1,q$ and $\infty$. Integrating over the Liouville zero mode, the correlator exhibits a pole whenever the neutrality condition upon charges of the vertex operators is satisfied \cite{Goulian1990}. The residue takes the form of coupled Selberg integrals similar to those appearing in the work of Dotsenko and Fateev \cite{Dotsenko1984a,*Dotsenko1984}. Upon the non-trivial assumption, investigated in \cite{Morozov2010}, that the prescription of the integration contours for screening charges is equivalent to the filling fraction condition in the standard large $N$ approach to matrix models, we are led to study the \bens\ partition function
\begin{equation}
\CZb(a,m_i,g,\b)=\int{\prod_{I=1}^N{e^{\frac{\sqrt{\b}}{g}V(\l_I)}d\l_I }\prod_{\superp{I,J=1}{I\neq J}}^{N}{\left(\l_I-\l_J\right)^{\b}}},
\end{equation}
with $g^2=-\e_1\e_2/4$ and $\b=-\e_2/\e_1$. The Penner type potential is reminiscent of the vertex operators. It introduces singularities at $x=0,1,q$ with residues being the rescaled Liouville charges $m_0,m_1,m_2$,
\begin{equation}\label{Penner}
V(x)=\left(m_0+(\e_1+\e_2)/2\right)\log x+m_1\log(x-1)+m_2\log(x-q).
\end{equation}
We choose here the potential corresponding to the model studied in \cite{Eguchi2010,Nishinaka2011,Bourgine2012} and mostly keep the notations of the reference \cite{Bourgine2012}. The dependence on the fourth charge $m_\infty$ appears through the neutrality condition
\begin{equation}\label{neutral}
\sum_{f=0,1,2,\infty}{m_f}+2\sqrt{\b}gN=0,
\end{equation}
where $N$ is the number of eigenvalues. According to the AGT correspondence \cite{Alday2009}, Liouville charges are related to the mass of hypermultiplets through 
\begin{equation}\label{def_mu}
\mu_\pm=m_2\pm m_0,\quad \tilde{\mu}_\pm=m_1\pm m_\infty.
\end{equation}

The NS limit is equivalent to a semi-classical limit for Liouville theory with heavy vertex operators. In this limit, the Liouville charges, as well as the background charge, have to be rescaled by a factor $\sim g$, and the bare charges tend to infinity as $g\to0$. The number of eigenvalues $N$ is also sent to infinity while $g\sqrt{\b}\sim\e_2$ goes to zero, such that the neutrality condition \ref{neutral} is satisfied. In this limit, the planar free energy of the \bens\ defined as
\begin{equation}
\CF_\b=\lim_{g\to0}{g^2\log \CZ_\b}
\end{equation}
is related to gauge theory free energy $\CF\full$ through
\begin{equation}\label{AGT_planar}
\CF\full=4\CF_\b+\left((m_0-m_2)^2+2\e(2m_0-\e)\right)\log q-2(m_1-\e)(m_2-\e)\log(q-1),
\end{equation}
up to a $q$- and $a$-independent additive constant \cite{Nishinaka2011}. As in the previous section, a density can be introduced, now associated to the eigenvalues $\l_I$,
\begin{equation}
\rho_\b(\l)=\lim_{g\to0}g\sqrt{\b}\la\sum_{I=1}^N\d(\l-\l_I)\ra.
\end{equation}
This density extremizes the Dyson effective action
\begin{equation}
\CA[\rho]=\int{V(\l)\rho(\l)d\l}+\int{\log|\l-\m|\rho(\l)\rho(\m)d\l d\m}+\dfrac{\e}2\int{\rho(\l)\log \rho(\l)d\l},
\end{equation}
where the last term in the RHS is a cut-off term coming from the measure and the regularization of the kernel at coincident eigenvalues \cite{Dyson1962}. The equation of motion derived from this action is equivalent to a Riccati equation for the shifted resolvent $H(x)$,
\begin{equation}\label{Ric}
H(x)^2+\e\p_xH(x)=T(x),\quad H(x)=2\int{\dfrac{\rho_\b(\l)d\l}{x-\l}}+V'(x).
\end{equation}
Apart from masses $m_f$ and equivariant deformation parameter $\e$, the function $T(x)$ also depends on a variable $E$. This variable is roughly equal to the gauge scalar amplitude $u_2=\la\tr\Phi^2\ra$, up to a gauge coupling dependent factor and a translation involving masses \cite{Eguchi2010a}. This amplitude is defined in \ref{def_amp} and parameterizes the gauge theory vacua. It can be evaluated in terms of the Coulomb branch vev $a$ using the $A$-cycle SW relation, equivalent to the filling fraction condition for the \bens. With a slight abuse of terminology, $T(x)$ is referred to as the ``stress energy tensor'' since it plays the role of this quantity in the Coulomb gas description \cite{Marshakov1991,*Kharchev1992,*Kostov1999}. It presents double poles on the sphere $\mathbb{CP}_1$ at $x=0,1,q,\infty$ and reads
\begin{equation}
T(x)=\dfrac{4p_4(x)}{x^2(x-1)^2(x-q)^2}+\dfrac{(q-1)E}{x(x-1)(x-q)},
\end{equation}
where $p_4(x)$ is a polynomial of degree four in $x$, quadratic in $\e$ and independent of $E$. Its expression will not be given here, but can be found in \cite{Bourgine2012} (equ. 2.20). In the SW limit, the Riccati equation \ref{Ric} reduces to an algebraic relation between $H$ and $x$ that defines the spectral curve. This spectral curve is a double cover of the sphere with four punctures, it has been identified with the SW curve in \cite{Eguchi2010,Eguchi2010a}. Introducing the wave function $\psi(x)$ as $H=\e\p\log\psi$, the Riccati equation transforms into a Schr\"odinger equation,
\begin{equation}\label{Schrod_gen}
(y^2-T(x))\psi(x)=0,\quad y=\e\p_x,\quad [y,x]=\e.
\end{equation}
Here, a notion of quantum curve emerges as the operator acting on $\psi$ is simply the spectral curve of the model at $\e=0$, with the function $H$  replaced by the operator $y$ \cite{Eynard2008a}. This Schr\"odinger equation can be derived directly within Liouville theory, without the \bens\ as an intermediate step \cite{Maruyoshi2010}.

SW relations for the DV \bens\ were derived in \cite{Bourgine2012} using the Dyson effective action. It was argued that because of poles at the branch points, it is necessary to subtract to the usual SW differential a function related to the cut off term of the action. However, to simplify the comparison with the gauge theory, we keep here the definition $dS=H(x)dx$ of the SW differential, and instead consider the $\e$-corrected relations
\begin{equation}\label{SW}
\dfrac{1}{2i\pi}\oint_\CA{dS}=a-\e/2,\qquad \oint_\CB{dS}=4\dfrac{\p\CF_\b}{\p a}+i\pi\e.
\end{equation}
These formulas are compatible with those found in \ref{SW_B} for $\CF\full$ under the AGT relation \ref{AGT_planar}.

\subsection{A quantum change of variable}\label{change_var}
We now express the AGT relation in the NS limit using our formalism. Similar approaches can be found in \cite{Zenkevich2011,*Mironov2012a,Mironov2012b}, where the connections with integrable models is discussed in more details. In particular, the quantum change of variable we present here has been found in \cite{Mironov2012b}. We reproduce it for the sake of completeness, and present a slightly different derivation. To keep expressions simple, we expose in this subsection only the massless case $\mu_\pm=\tmu_\pm=0$. The massive case is treated in appendix \refOld{App_fullmass} and requires extra care to cancel poles. Our starting point is the Schr\"odinger equation \ref{Schrod_gen},
\begin{equation}
\la x\Big|y^2+\dfrac{(\e/2)^2}{x^2}-\dfrac{(q-1)E}{x(x-1)(x-q)}\Big|\psi\ra=0.
\end{equation}
In order to emphasize the fact that we are performing a quantum computation, we used a bracket notation, and commutation of variables must be performed cautiously. The first step is to multiply on the left by the product $x(x-1)(x-q)$ in order to eliminate the poles at $x=1,q$ and reduce the order of the pole at $x=0$. Then, we introduce the quantum change of variable $z=xy-\e/2$. Expressing $y$ in terms of $z$, we get
\begin{equation}
\la x|(x-1)(x-q)x^{-1}z^2-(q-1)E|\psi\ra=0.
\end{equation}
Classically, this equation is simply the Seiberg-Witten curve \cite{Seiberg1994} in the Gaiotto coordinates \cite{Gaiotto2009a}.

The commutation relation between $x$ and $z$ arises from the canonical commutator between $x$ and $y$. Preservation of commutation relations is required to perform a canonical change of variable $(x,y)\to(z,p)$ with $p=\e\p_z$. It is noted that this commutation relation remains unchanged when $z$ is shifted by an arbitrary function of $x$. This property is exploited in the massive case to further define $z=xy+s(x)$ with an appropriate function $s(x)$. For the representation of the operator $x$ in the basis $|z>$ to be compatible with the commutator $[x,z]=-\e x$, $x$ must act as a shift operator on functions of $z$, with a possible multiplicative factor $r$. Choosing
\begin{equation}\label{expr_x}
x=r(z)e^{-p},\quad x^{-1}=e^pr(z)^{-1},\quad\text{with}\quad r(z)=qz^2,
\end{equation}
the Schr\"odinger equation becomes a difference equation,
\begin{equation}
\la x|q M(z) e^{-\e\p_z}-(1+q)P(z)+e^{\e\p_z}|\psi\ra=0,
\end{equation}
which is precisely the Baxter TQ relation \ref{BaxterTQ}. We also recover the proper expression for the mass polynomial $M(z)$ as given in \ref{def_M} with $\mu_\pm=\mut_\pm=0$, and derived the expression of the monic polynomial $P(z)$ of degree two. In the massive case, the functions $s(x)$ and $r(z)$ involved in the change of variable acquire a mass dependence,
\begin{equation}\label{sr}
s(x)=\mu_+-xV'(x),\quad r(z)=q(z-\mu_+)(z-\mu_-),
\end{equation}
where the \bens\ potential $V(x)$ is defined in \ref{Penner}. This change of variable is not the only one possible. Other choices include: sign flip of $y$ in the definition of $z$, translation of $s(x)$ by a constant, flip of the masses $m_{1,2}\to\e-m_{1,2}$ appearing in the residues of $s(x)$ at $x=1$ and $x=q$,... These different choices lead to different mass polynomials. However, it can be shown that they describe the same theory by exploiting various invariance such as the sign flip of $a$, or the exchange of fundamental and anti-fundamental hypermultiplets. Although probably worth of investigation, the inventory of all possible choices will not be done here, and we focus on the one given in \ref{sr}.\\

\subsection{Wave functions transformation}
From the analysis performed in the previous subsection, we conclude that the wave functions $\psi(x)$ and $\hpsi(z)$ represent the same state $|\psi>$ expressed either in the $|x>$ or $|z>$ canonical bases. It means that the two wave functions are related through an integral transform. The corresponding kernels $<x|z>$ and $<z|x>$ solve the constraints associated to the representation of the operator $z$ and $x$ in the $|x>$ and $|z>$ basis respectively. They also have to satisfy a formal inversion property. Like the function $\hpsi$, these kernels are determined up to an $\e$-periodic $z$-factor. This degree of freedom arises on the gauge theory side because the mapping $(y,x)\to(p,z)$ is not one to one, but leave the possibility to shift $p$ by a multiple of $2i\pi$. Thus, we have shown that the AGT relation takes the form of a Mellin transform between wave functions,
\begin{equation}\label{Mellin}
\hpsi(z)=\sigma(z)\int_0^\infty{dx x^{\frac{\mu_+-z}{\e}-1}e^{-\frac1\e V(x)}\psi(x)},\quad \psi(x)=\dfrac{e^{\frac1\e V(x)}}{2i\pi\e}\int_{c-i\e\infty}^{c+i\e\infty}{dz x^{\frac{z-\mu_+}{\e}}\sigma(z)^{-1}\hpsi(z)}.
\end{equation}
The appearance of the Mellin transform was expected here since when $r(z)=1$ the momentum $p$ is simply minus the logarithm of the coordinate $x$, and the change of canonical basis between $p$ and $z$ is known to be characterized by an exponential kernel. It is worth noticing that in \ref{Mellin} the potential term compensates the translation by $V'$ in the definition \ref{Ric} of the resolvent $H$. This shift also appears when the wave function $\psi$ is expressed as the \bens\ correlator
\begin{equation}
\psi(x)=e^{\frac1{2\hbar} V(x)}\lim_{g\to0}\la\prod_{I=1}^{N}(x-\l_I)\ra.
\end{equation}
The function $\s(z)$ plays the role of a potential term but for the function $\hpsi(z)$, it is a solution of the difference equation $r(z)=\s(z)/\s(z-\e)$. Up to an irrelevant $\e$-periodic factor, it can be taken to be 
\begin{equation}
\s(z)\equiv(q\e^2)^{z/\e}\G\left[1+(z-\mu_+)/\e\right]\G\left[1+(z-\mu_-)/\e\right].
\end{equation}
Replacing $\hpsi\to\s^{-1}\hpsi$ boils down to multiply $\o$ by $r$ which corresponds to shift the density $f''(t)$ by mass delta functions, as can be seen comparing \ref{def_o} with
\begin{equation}
\log r(z)=\sum_\pm\int_\mathbb{R}{\d(t-\mu_\pm)\log(z-t)dt}+\log q.
\end{equation}
As for the $|x>$ space, there is a choice of coordinate, through $r(z)$, such that the difference equation over $\o$ becomes trivial. However, finding this coordinate is of the same difficulty as solving the full equation. The convergence of the first integral transform in \ref{Mellin} depends on the behavior of the function $\psi(x)$ at zero and infinity. The Schr\"odinger equation \ref{Schrod_gen} has two linearly independent solutions $\psi_\pm$ with a different asymptotic,
\begin{equation}
\psi_\pm(x)e^{-\frac1\e V(x)}\sim x^{-\frac1\e(\mu_++\tmu_\pm)},\quad x\to\infty.
\end{equation}
The physical solution, or more precisely the wave function reproducing the resolvent on the physical sheet, is $\psi_+$. As $x\to 0$, the two solutions behave as
\begin{equation}
\psi_+(x)e^{-\frac1\e V(x)}\sim 1,\quad \psi_-(x)e^{-\frac1\e V(x)}\sim x^{-\frac1\e(\mu_+-\mu_-)},\quad x\to 0.
\end{equation}
Supposing $\Re\e>0$, the integral over $x$ converges when $\mu_\pm>z>-\tmu_\pm$ for respectively the wave functions $\psi_\pm(x)$. Values of $\hpsi(z)$ for $z$ outside this interval must be obtained from analytical continuation, which implies a deformation of the integration contour in \ref{Mellin}. The transformation is worked out in details in appendix \refOld{App_cv} for the simple case of a stress-energy tensor with only one double pole.\\

\subsection{Mapping between SW differential forms}
In the limit $\e\to0$, the inverse Mellin transform can be approximated by the method of steepest descent. In general, employing this saddle point technique on a Laplace transform leads to a Legendre transform at the first order of approximation \cite{Zia2009}. For this purpose, we introduce a notation for the phase of wave functions,
\begin{equation}
\psi(x)=e^{\frac{1}{\e}\phi(x)},\qquad \hpsi(z)=e^{-\frac1{\e}\hphi(z)}.
\end{equation}
Using the CFT methods for matrix models \cite{Marshakov1991}, with a natural generalization to \bens s by turning on a background charge \cite{Itoyama2009}, the function $\phi(x)$ is identified with a Coulomb gas field. This field is roughly the chiral part of the Liouville field after integration over the zero mode and a Wick rotation. As $\e\to 0$, the transformation
\begin{equation}
e^{\frac1\e(\phi(x)-V(x))}=\dfrac1{2i\pi\e}\int_{c-i\e\infty}^{c+i\e\infty}{dz e^{\frac1\e\left[(z-\mu_+)\log x-\hphi(z)-\e\log\s(z)\right]}}
\end{equation}
reduces to a Legendre transform between $\phi$ and $\hphi$,
\begin{equation}\label{Legendre}
\phi(x)-V(x)=(z-\mu_+)\log x-(\hphi(z)+\e\log\s(z)).
\end{equation}
The mapping $x(z)$ is obtained as a solution of the saddle point equation
\begin{equation}\label{saddle_L}
\log x=-\e\p\log(\hpsi(z)/\s(z)).
\end{equation}
In the SW limit, this is equivalent to set $x=r(z)\o_0(z)$. Up to the function $r$ which can be absorbed in a redefinition of $\o$, we recover the classical change of variable performed by Gaiotto in \cite{Gaiotto2009a}: the original differential form $zd\o_0/\o_0$ is replaced by $ydx$ with $x=\o_0$ and $z=\o_0 y=xy$. Thus, the transformation $(p,z)\to(y,x)$ can be interpreted as the quantized form of the Gaiotto change of variables.

On the Liouville side, the SW differential is simply $dS=d\phi$. It is actually more convenient to write it as $dS=x\p_x\phi(x)d\log x$. The Legendre transform implies 
\begin{equation}\label{dphi}
\p_{\log x}\phi(x)=xV'(x)+(z-\mu_+),
\end{equation}
and $\log x$ is given by \ref{saddle_L}. The terms containing $V'$, $\mu_+$ or $\s$ generate no contribution to $A$- or $B$-cycle integrals and can be neglected in the definition of the SW differential. They will however modify the pole structure of the $1$-form, replacing the poles at $x=0,1,q,\infty$ by poles at $z=\mu_\pm,\tmu_\pm$. We deduce the expression of $dS$ in the basis $|z>$ at the first order, upon integration by parts, $dS=-d\hphi$, i.e. $dS=zd\log\o_0$. We recover here the expression \ref{def_SW2} of the previous section, up to a sign. This sign seems to indicate that the physical and non-physical sheets are interchanged under the transformation \ref{Mellin}: the physical wave function $\psi_+$ gives rise to $\hpsi_+$ which produces the values of the resolvent on the non-physical sheet ($\o_+$) through the relation \ref{ratio_psi}.

At the next order, we need to take into account the square root of the Hessian in the evaluation of the integral, and the Legendre transform \ref{Legendre} get $\e$-corrections of the form,
\begin{equation}
\phi(x)-V(x)=(z-\mu_+)\log x-(\hphi(z)+\e\log\s(z))-(\e/2)\log(\p^2\log(\hpsi(z)/\s(z)))
\end{equation}
However, the saddle point equation \ref{saddle_L} and the relation \ref{dphi} still holds. It implies that the SW differential is again given by the expression $dS=-d\hphi$, in agreement with the results from the gauge theory side. Higher order corrections to the saddle point approximation do not modify the equation \ref{dphi}, and the equality $dS=-d\hphi$ actually hold at all order in $\e$. It can be seen as a proof of the matrix model version of the AGT conjecture in the NS limit and up to an $a$-independent term. To go further and prove the AGT relation \ref{AGT_planar} including the $a$-independent constant requires a better understanding of the transformation for the densities and actions. Finding the correct change of variable for the functional integral of the collective field theories would be a first step in the derivation of a matrix model transformation between the \bens\ and the gauge theory partition function.

As a final remark, let us stress that the study presented in this paper is restricted to the case of a single $SU(N_c)$ gauge group. It would be tempting to extend the picture to the whole set of quiver theories, as was done in \cite{Nekrasov2012} for the SW limit, lately extended to the NS background \cite{Fucito2012}. To each node of the quiver diagram should be associated a different density $\rho$, or limit shape $f$. On the other hand, densities introduced in the \bens\ representing Toda correlators are in correspondence with the number of fields, i.e. the rank $N_c-1$ of a single group. The generalization of the transformation between wave functions associated to each density is still an open problem.

\section*{Acknowledgments}
I would like to thank Y. Matsuo for instructive discussions and for sharing his preliminary results. It is also a pleasure to acknowledge the warm hospitality of Yukawa institute during the workshop ``Gauge/Gravity Duality'' where part of this work was done.

\appendix
\section{Definitions and properties of the functions $\g$}\label{AppA}
The function $\g_{\e_1,\e_2}$ is defined as the logarithm of the Barnes double gamma function \cite{Alday2009}. It is invariant under the exchange of the parameters $\e_1$ and $\e_2$, and can be expressed as the following integral,
\begin{equation}
\g_{\e_1,\e_2}(x)=\e_1\e_2\log\G_2(x|\e_1,\e_2)=\e_1\e_2\left.\dfrac{d}{ds}\right|_{s=0}\dfrac1{\G(s)}\int_0^\infty{\dfrac{dt}{t}t^s\dfrac{e^{-tx}}{(1-e^{-\e_1t})(1-e^{-\e_2t})}}.
\end{equation}
We included here the factors $\e_1\e_2$ to have well defined limits
\begin{equation}\label{def_ge}
\g_{\e}(x)=\lim_{\e_2\to0}\g_{\e,\e_2}(x)=\e\left.\dfrac{d}{ds}\right|_{s=0}\dfrac1{\G(s)}\int_0^\infty{\dfrac{dt}{t^2}t^s\dfrac{e^{-tx}}{(1-e^{-\e t})}},\quad \g(x)=\lim_{\e\to0}\g_{\e}(x)=-\dfrac{x^2}{2}(\log x-3/2).
\end{equation}
The $\g$-functions satisfy the difference equations
\begin{equation}\label{shift}
(1-e^{\e_2\p_x})\g_{\e_1,\e_2}(x)=-\e_2\g_{\e_1}'(x),\quad (1-e^{\e\p_x})\g_\e(x)=-\e\g'(x).
\end{equation}
The expansion at large $x$ for $|\arg(x)|<\pi$ of the function $\g_{\e_1,\e_2}(x)$ is \cite{Spreafico}
\begin{equation}
\g_{\e_1,\e_2}(x)=-\hf x^2(\log x-3/2)+\hf(\e_1+\e_2)x(\log x-1)-\dfrac14\left(\e_1\e_2+\dfrac13(\e_1^2+\e_2^2)\right)\log x+O(1)
\end{equation}
which implies
\begin{equation}
\g_{\e}(x)=-\hf x^2(\log x-3/2)+\hf\e x(\log x-1)-\dfrac1{12}\e^2\log x+O(1).
\end{equation}
This expansion is consistent with the limit $\e\to0$ reproducing $\g(x)$. The integral expression for $\g_\e^{(3)}$ is not singular at $t=0$ and can be computed without the $\zeta_2$-regularization. It is proportional to the derivative of the digamma function with argument $x/\e$. Integrating twice, and comparing the asymptotic expansion of $\g'_\e$ with $\log\G$, we get
\begin{equation}
\g'_\e(x)=-\e\log\G[x/\e]-x\log\e+\dfrac{\e}{2}\log(2\pi\e).
\end{equation}

Instead of $\g_\e$, it reveals simpler to study directly the function $\tg_\e$ defined through its relation to the function $G^{II}$ given in \ref{def_GII},
\begin{equation}\label{G_tge}
\e G^{II}(x)=2(1-e^{-\e\p_x})(1-e^{\e\p_x})\tg_\e(x).
\end{equation}
This definition allows for an arbitrary shift of $\tg_\e$ by an affine function of $x$. Exploiting the relation \ref{rel_Gg} between $G^{II}$ and $\g_\e$, we choose
\begin{equation}
\tg_\e(x)=\hf(1+e^{\e\p_x})\g_\e(x)-\dfrac{\e}2x\log 2+\text{cste},
\end{equation}
where the constant is left undefined as it plays no role in our discussion. Explicitly,
\begin{equation}
\tg_\e'(x)=-\e\log\G[x/\e]-(\e/2)\log x-x\log \e+(\e/2)\log(\pi\e).
\end{equation}
From the property
\begin{equation}
\G[x]\G[-x]=-\dfrac{\pi}{x\sin\pi x},
\end{equation}
we derive the relation
\begin{equation}
\tg_\e'(-x)=-\tg_\e'(x)+\e l(x/\e),\quad\text{with}\quad l(x)=\log|\sin\pi x|,
\end{equation}
up to shifts of $\pm i\pi\e/2$ and $\pm i\pi\e$, depending on the position of $x$ with respect to the branch cut of the logarithm, and on the sign of the sine.

\section{Analysis of a simple difference equation}\label{AppB}
In this appendix, we study the difference equation
\begin{equation}\label{equ_rho}
\left(1-e^{-\e\p_t}\right)\rho(t)=\d(t-a).
\end{equation}
This equation admits solutions defined up to a translation by an $\e$-periodic function. It has a formal solution in the form of an infinite series of $\d$ functions. However, we instead perform here a perturbative analysis in $\e$. At the first order, we find
\begin{equation}
\e\rho_0'(t)=\d(t-a)\implies \rho_0(t)=\dfrac1{\e}\Theta(t-a)+\g_0
\end{equation}
where $\Theta(t)$ is the Heaviside function, and $\g_0$ a constant of integration. This constant can be fixed by requiring that $\rho_0$ has a finite norm using the principal value regularization. We find the value $\g_0=-1/2\e$ which gives
\begin{equation}
\rho_0(t)=\dfrac1{2\e}\sign(t-a),\quad \int_{\mathbb{R}}{\rho_0(t)dt}=-\dfrac{a}{\e}.
\end{equation}
At the next order, decomposing $\rho=\rho_0+\e\rho_1$, we find
\begin{equation}
\rho_1(t)=\dfrac1{2\e}\d(t-a).
\end{equation}
We choose here the integration constant to be zero in order to keep the norm of $\rho$ finite. We shall take the same choice for all higher orders. Up to the fifth order, we have
\begin{equation}
\rho(t)=\dfrac1{2\e}\sign(t-a)+\hf\d(t-a)+\dfrac1{12}\e\d'(t-a)-\dfrac1{6!}\e^3\d^{(3)}(t-a)+O(\e^5).
\end{equation}
At the order $O(\e^l)$, the density $\rho_l$ is given by the sum over the $(k-l-1)$th derivative of $\rho_k$, which are proportional to $\d^{(l-1)}(t-a)$ by recursion. It implies that the density $\rho$ is expressed as the infinite series
\begin{equation}\label{rho_serie}
\rho(t)=\dfrac1{2\e}\sign(t-a)+\sum_{l=1}^\infty{\e^{l-1}c_l\d^{(l-1)}(t-a)}.
\end{equation}
The coefficients $c_l$ can be obtained from the small $x$ expansion of the generating function $x/(1-e^{-x})$. We find $c_0=1$, $c_1=1/2$ and for $l>1$, $c_l$ is vanishing when $l$ is odd, and is expressed in terms of the Bernouilli numbers $c_l=B_l/l!$ when $l$ is even. It follows from the expression \ref{rho_serie} that the moments
\begin{equation}
\mu_k=\int_{-\L}^{\L}{t^k\rho(t)dt}
\end{equation}
can be computed using only the $k+2$ first order terms $\rho_0,\cdots,\rho_{k+1}$ in the $\e$-expansion of $\rho$. In the definition of the moments $\mu_k$, we used a principal value regularization, and $\L$ should be sent to infinity at the end of the calculations. Using this perturbative approach, we find the following values of the moments,
\begin{align}
\begin{split}
&\e\mu_0=-a+\e/2,\quad \e\mu_1=\dfrac{\L^2}{2}-\hf\left(a^2-a\e+\e^2/6\right),\\
&\e\mu_k=\dfrac{1+(-1)^{k+1}}{2(k+1)}\L^{k+1}-\dfrac1{(k+1)}a^{k+1}+\e\sum_{l=0}^k{c_{l+1}\dfrac{k!}{(k-l)!}(-\e)^la^{k-l}}.
\end{split}
\end{align}
Cut-off contributions only appear for even moments.

There exists a different approach to compute the moments based on the following formula,
\begin{equation}
\int_{-\L}^{\L}{\rho(t)\left(1-e^{\e\p_t}\right)t^kdt}=a^k-\dfrac1{2(k+1)\e}\left[(\e-\L)^{k+1}+(\e+\L)^{k+1}-\L^{k+1}-(-\L)^{k+1}\right].
\end{equation}
This expression is established using a change of variable over $t$ to transpose the shift operator over the density $\rho$ in order introduce the difference equation \ref{equ_rho}. The second term is a boundary term coming from the finite bound of integration used in the regularization. In the process, the delta functions (and derivatives) evaluated at $\pm\L=a$ and $\pm\L=a-\e$ are discarded since $\L$ is much larger than $a$. Note that the boundary term is independent of $a$, and will be canceled when considering the difference between $\rho\ol$ and $\rho\ct $. Such a boundary term typically occurs when the density $\rho$ is replaced by the shape function $f$. At first orders, we have
\begin{equation}
\int_{-\L}^{\L}{\rho(t)\left(1-e^{\e\p_t}\right)dt}=0,\quad \int_{-\L}^{\L}{\rho(t)\left(1-e^{\e\p_t}\right)tdt}=a-\e/2,\quad \int_{-\L}^{\L}{\rho(t)\left(1-e^{\e\p_t}\right)t^2dt}=a^2-\L^2-\e^2/3.
\end{equation}
In particular, one can check the compatibility between the two methods,
\begin{equation}
\int_{-\L}^{\L}{\rho(t)\left(1-e^{\e\p_t}\right)tdt}=-\e\mu_0,\quad \int_{-\L}^{\L}{\rho(t)\left(1-e^{\e\p_t}\right)(t^2-\e t)dt}=-2\e\mu_1.
\end{equation}

To make contact with the work described in \cite{Ferrari2012a}, we now employ a Fourier transform to solve \ref{equ_rho}. In the Fourier space,
\begin{equation}
\hat\rho(\t)=\int_\mathbb{R}{\rho(t)e^{it\t}dt}=-\dfrac{1}{2i\sin(\e\t/2)}e^{i\t(a-\e/2)}.
\end{equation}
By linearity, it is sufficient to take the sum over the color index to solve the equation \ref{def_rp} for $\rho\ol$. The expression for the Fourier transformed of the primitive $\brho\ol$ is obtained after dividing $\hat\rho\ol$ by $i\t$, and writes
\begin{equation}
\hat\brho\ol(\t)=\dfrac{1}{2\t\sin(\e\t/2)}\sum_{l=1}^{N_c}{e^{i\t(a_l-\e/2)}}.
\end{equation}
We recover here the expression of $\tilde\rho_{Q_0}=-\hat\brho\ol$ found in \cite{Ferrari2012a} upon multiplication above and below by $\cos(\e\t/2)$. The pole at $\t=0$ is related to the norm of $\rho\ol(\t)$ which is divergent. However, this infinity is compensated by the presence of $\rho\ct $ in $\rho$ and we do not see any reason here to impose a finite norm, in contrast with the claim of \cite{Ferrari2012a}.

\section{Derivation of the change of variable in the massive case}\label{App_fullmass}
Starting from the Schr\"odinger equation \ref{Schrod_gen}, we again multiply on the left by $x(x-1)(x-q)$. It eliminates the double poles at $x=0,1,q$, leaving only single poles. The remaining poles at $x=1$ and $x=q$ are eliminated by a good choice of the function $s(x)$ in the change of variable $z=xy+s(x)$. With this change of variable, we have
\begin{equation}
x^2y^2=z^2-2zs(x)+\e xs'(x)+s(x)^2-\e(z-s(x)).
\end{equation}
Using commutation relations, we push all $x$-dependence to the right, and write the equation as
\begin{equation}
\la x\Big|Z(x)+R(x)x^{-1}-(q-1)E\Big|\psi\ra=0,
\end{equation}
In this expression, terms that may present poles at $x=1$ and $x=q$ have been gathered in the operator $R(x)$ which is $z$-independent,
\begin{equation}
R(x)=2\e s(x)(x^2-q)+(x-1)(x-q)\left[s(x)^2+\e xs'(x)\right]-\dfrac{4p_4(x)}{(x-1)(x-q)}.
\end{equation}
The remaining $z$-dependent term is free of singularities at $x=1,q$ provided that $s(x)$ has no more than first order poles,
\begin{equation}
Z(x)=(z-\e)(z-2\e)x-(1+q)z(z-\e)+qz(z+\e)x^{-1}+(\e-2z)s(x)x^{-1}(x-1)(x-q).
\end{equation}
The function $s(x)$ is determined from the requirement that $R(x)$ is a polynomial in $x$ of degree at most two. In particular, poles at $x=1,q$ must vanish. We propose the ansatz
\begin{equation}
s(x)=\a+\dfrac{s_1}{x-1}+\dfrac{s_q}{x-q}.
\end{equation}
To fulfill the requirement of vanishing residue at $x=1$ and $x=q$ on $R(x)$ we take $s_1=-m_1$ and $s_q=-qm_2$. It is worth noticing that the constants $s_1$ and $s_q$ are solutions of a quadratic equation and should a priori be infinite series in $\e$. Surprisingly, the solution we find has no $\e$-corrections. Under this choice of $s(x)$, and regrouping the terms according to their $x$-dependence, we find
\begin{equation}
\la x\Big|q\dfrac{M(z)}{r(z)}x-(1+q)P(z)+r(z+\e) x^{-1}\Big|\psi\ra=0,
\end{equation}
which is exactly the form of the Baxter TQ relation \ref{BaxterTQ}. In general $r(z)$ depends on three variables $m_0$, $m_1$ and $m_2$, which leads to an incorrect form of the function $M(z)$. To reproduce the correct mass polynomial as given in \ref{def_M}, we have to set $\a=-m_1-\e/2$ in $s(x)$ which suppress the dependence in the mass $m_1$ of $r(z)$. Then, the zeros of $r(z)$ coincide with the values of $\mu_+$ and $\mu_-$ given in \ref{def_mu}, and we recover the expression \ref{def_M} for $M(z)$. Finally, the polynomial $P(z)$ is indeed monic and of degree two,\footnote{Another solution is possible, $s_1=m_1-\e$ and $s_q=q(m_q-\e)$. In this case, to identify properly the mass polynomial $M(z)$, we need to take into account the action of the Weyl group of the $SU(2)$ gauge symmetry that flips $a\to-a$, $m_0\to-m_0$, $m_\infty\to-m_\infty$, $m_1\to\e-m_1$ and $m_2\to\e-m_2$ \cite{Alday2009}. It leads to define
\begin{equation}
\mu_\pm=\e-m_2\pm m_0,\quad \tilde{\mu}_\pm=\e-m_1\pm m_\infty,
\end{equation}
and we recover the correct Baxter relation with a different polynomial $P(z)$.}
\begin{align}
\begin{split}
P(z)&=z^2-(e_1+e_2)z+e_1e_2,\qquad e_1+e_2=\dfrac{2q}{1+q}(m_2-m_1),\\
(1+q)e_1e_2&=-(1-q)E-q\left(m_\infty^2-3m_2^2+2(m_1+\e)m_2-m_1^2-m_0^2\right)-2m_2(m_2-\e)-2m_0^2.
\end{split}
\end{align}
Gathering all the constraints, the function $s(x)$ can be expressed with the help of the \bens\ potential as in \ref{sr}. In the massless limit, it reduces to $s(x)=-\e/2$.

\section{Study of the Mellin transform between wave functions}\label{App_cv}
In this appendix, we examine the quantum change of variable $z=xy-\e/2$ for the wave functions $\psi_{\a,\g}=x^{\a/\e} e^{-\g x/\e}$. We suppose here $\g>0$, in the case $\g<0$ the integration range $[0,+\infty[$ of the Mellin transformed should be replaced by $]-\infty,0]$. These wave functions satisfy a Schr\"odinger equation with a potential having a double pole at $x=0$,
\begin{equation}
y^2\psi_{\a\g}(x)=T(x)\psi_{\a\g}(x),\quad T(x)=\a(\a-\e)\dfrac1{x^2}-2\a\g\dfrac1x+\g^2.
\end{equation}
Such a potential would appear in the study of a sphere with a single puncture at $x=0$. Performing the quantum change of variable over this Schr\"odinger equation with $r(z)=z^2-(\a-\e/2)^2$, we get the difference equation
\begin{equation}\label{equa_diff}
\g^2M(z)\hpsi_{\a\g}(z-\e)-2\a\g\hpsi_{\a\g}(z)-\hpsi_{\a\g}(z+\e)=0,\quad M(z)=r(z).
\end{equation}
It is easy to work out the transformation of $\psi_{\a\g}$, we get
\begin{equation}
\hpsi_{\a\g}(z)=-\dfrac{\pi}{\cos((z-\a)/\e)}\g^{(z-\a)/\e+1/2}\e^{(z+\a)/\e-1/2}\G[1/2+(z+\a)/\e].
\end{equation}
It is also trivial to check that this wave function indeed satisfies the difference equation \ref{equa_diff}.

It was shown in \cite{Bourgine2012} that the definition of the SW differential coincides with the coordinate transformation that trivializes the stress energy tensor of the Coulomb gas representation. In these new coordinates, that for simplicity we still denote $x$ and $y$, we have $dS=dx$ and $T(x)=1$. This problem corresponds to the case we just treated, with the special values $\a=0$ and $\g=1$. We note that the difference equation simplifies in this case since the term proportional to $\hpsi(z)$ disappear. Under the choice of coordinates $r(z)=z+\e/2$, the factor $M(z)$ disappear and the trivial wave function $\hpsi(z)=1$ solves the equation. This choice of coordinate modifies the integration kernel of the Mellin transform through the function $\s(z)$ and indeed produces a trivial, i.e. $\e$-periodic, wave function.

\bibliographystyle{unsrt}

\end{document}